\documentclass[fleqn,usenatbib]{mnras}

\usepackage{newtxtext,newtxmath}

\usepackage[T1]{fontenc}
\usepackage{ae,aecompl}

\usepackage{graphicx}	
\usepackage{amsmath}	
\usepackage{amssymb}	

\newcommand{\kms}{\ifmmode  \,\rm km\,s^{-1} \else $\,\rm km\,s^{-1}  $ \fi}
\newcommand{\Mpc}{\ifmmode  {\rm~Mpc}  \else ${\rm~Mpc}$\fi}
\newcommand{\kpc}{\ifmmode  {\rm~kpc}  \else ${\rm~kpc}$\fi}
\newcommand{\Gyr}{\ifmmode  {\rm~Gyr}  \else ${\rm~Gyr}$\fi}
\newcommand{\Msun}{\ifmmode {\rm M}_{\odot} \else ${\rm M}_{\odot}$ \fi}
\newcommand{\kmsMpc}{\ifmmode  \,\rm km\,s^{-1}\,Mpc^{-1} \else $\,\rm km\,s^{-1}\,Mpc^{-1}  $ \fi }
\newcommand{\Omegam}{\ifmmode \Omega_{\rm m} \else $\Omega_{\rm m}$\fi} 
\newcommand{\OmegaL}{\ifmmode \Omega_{\rm \Lambda} \else $\Omega_{\rm \Lambda}$\fi} 
\newcommand{\Omegab}{\ifmmode \Omega_{\rm b} \else $\Omega_{\rm b}$\fi} 
\newlength{\fullwidth}
\newlength{\halfwidth}
\title[Empirical constraints on ETG formation]
{Empirical constraints on the formation of early-type galaxies}

\author[B. P. Moster et al.]{
Benjamin P. Moster,$^{1,2}$\thanks{E-mail: moster@usm.lmu.de (BPM)}
Thorsten Naab,$^{2}$
Simon D. M. White$^{2}$
\\
$^{1}$Universit\"ats-Sternwarte, Ludwig-Maximilians-Universit\"at M\"unchen, Scheinerstr. 1, 81679 M\"unchen, Germany\\
$^{2}$Max-Planck-Institut f\"ur Astrophysik, Karl-Schwarzschild Stra\ss e 1, 85748 Garching, Germany\\
}

\date{Accepted XXX. Received YYY; in original form ZZZ}

\pubyear{2019}

\begin{document}
\label{firstpage}
\pagerange{\pageref{firstpage}--\pageref{lastpage}}
\maketitle


\begin{abstract}
We present constraints on the formation and evolution of early-type galaxies (ETGs) with the empirical model {\sc emerge}. The parameters of this model are adjusted so that it reproduces the evolution of stellar mass functions, specific star formation rates, and cosmic star formation rates since $z\approx10$ as well as `quenched' galaxy fractions and correlation functions. We find that at fixed halo mass present-day ETGs are more massive than late-type galaxies, whereas at fixed stellar mass ETGs populate more massive halos in agreement with lensing results. This effect naturally results from the shape and scatter of the stellar-to-halo mass relation and the galaxy formation histories. The ETG stellar mass assembly is dominated by `in-situ' star formation below a stellar mass of $3\times10^{11}\Msun$ and by merging and accretion of `ex-situ' formed stars at higher mass. The mass dependence is in tension with current cosmological simulations. Lower mass ETGs show extended star formation towards low redshift in agreement with recent estimates from IFU surveys. All ETGs have main progenitors on the `main sequence of star formation' with the `red sequence' appearing at $z \approx 2$. Above this redshift, over 95 per cent of the ETG progenitors are star-forming. More than 90 per cent of $z \approx 2$ `main sequence' galaxies with $m_* > 10^{10}\Msun$ evolve into present-day ETGs. Above redshift 6, more than 80 per cent of the observed stellar mass functions above $10^{9}\Msun$ can be accounted for by ETG progenitors with $m_* > 10^{10}\Msun$. This implies that current and future high redshift observations mainly probe the birth of present-day ETGs. The source code and documentation of \textsc{emerge} are available at \href{https://github.com/bmoster/emerge}{github.com/bmoster/emerge}.
\end{abstract}





\section{Introduction} \label{sec:intro}

In the $\Lambda$CDM framework, the Universe consists of dark energy that can be described by a cosmological constant $\Lambda$, dynamically cold dark matter (CDM), and a small fraction of baryonic matter. In this scenario, structure formation proceeds hierarchically through gravitationally driven collapse and the merging of smaller structures. In the standard picture of galaxy formation, gas can cool within virialised dark matter haloes and form stars at their centres \citep{White:1978aa, Fall:1980aa, Blumenthal:1984aa}. Once a halo falls into a more massive halo and merges, its galaxy becomes a satellite and may eventually merge with the central galaxy of the main halo impacting its properties. In this way, galaxies can grow both by converting their gas into stars (in-situ), and by accreting stars from merging satellites (ex-situ). As the properties of the haloes determine how much gas is available for star formation and how efficiently this conversion can proceed, the formation histories of the haloes directly impact the formation histories of their galaxies. Consequently, the relation between the stellar mass of a galaxy and the mass of the halo in which it is embedded depends on the average formation time of the halo.

Observed galaxies can be arranged into a continuous sequence of types, as originally introduced by \citet{Hubble:1926aa}. Early type galaxies (ETGs) on the one end typically have elliptical shapes with little internal structure, while late type galaxies (LTGs) on the other end show diverse substructure such as discs with spiral arms. ETGs are the non-star-forming (or `quenched'), dynamically hot and massive systems of the present-day galaxy population, whereas LTGs tend to be actively star-forming, dynamically cold and less massive. In the current picture, massive galaxies are more likely to have violent merger histories leading to the elliptical morphologies of ETGs. Weak gravitational lensing studies show that ETGs tend to live in more massive haloes than their LTG counterparts of the same stellar mass \citep{Mandelbaum:2006aa,Mandelbaum:2016aa}. This is further evidence for different formation histories of ETGs and LTGs. All archaeological stellar population studies indicate that the most massive present-day ETGs with $\log(m_*/\Msun)\gtrsim11$ formed the vast majority of their stars at high redshift ($z \gtrsim ~ 2$) on short timescales of the order of $1\Gyr$ \citep{Gavazzi:2002aa,Heavens:2004aa,Thomas:2005aa,Thomas:2010aa,McDermid:2015aa}. For lower mass ETGs there is a trend towards lower formation redshifts and longer formation time scales. In contrast, it has been shown that low-mass dark matter haloes assemble early, and massive haloes assemble late \citep{Lacey:1993aa,Sheth:2004aa,Boylan-Kolchin:2009aa}. This apparent discrepancy has been termed `anti-hierarchical' galaxy formation or `downsizing' \citep{Cowie:1996aa}. The dichotomy has been addressed by \citet{DeLucia:2006aa}, who show how this effect naturally arises in the hierarchical paradigm. While the assembly of massive galaxies is hierarchical, AGN feedback leads to low conversion efficiencies in massive haloes, such that all stars formed at early times, before the onset of AGN feedback. Further, in a hierarchical cosmological model the formation time and the assembly time of stars in galaxies can be decoupled, as galaxies also grow by merging \citep[e.g.][]{Oser:2010aa}. Since most of the mergers that bring in a significant amount of stellar mass happen at late cosmic  times, the assembly times are typically much later than the formation times of the stars.

The cosmological formation and assembly histories of ETGs have been studied with various theoretical approaches. The most common methods employ a specified set of physical processes for the baryons to evolve an initial distribution of dark matter and gas. In hydrodynamic simulations the gas component is followed along with the dark matter, and the physical processes such as gas cooling, star formation, and several feedback mechanisms are modelled on a chosen physical resolution scale \citep{Springel:2005ac,Teyssier:2013aa,Hopkins:2014aa,Wang:2015aa}. Semi-analytic models (SAMs) separate the formation of galaxies from the formation of  gravitationally  bound structures by post-processing halo merger trees with physically motivated recipes \citep{White:1991aa,Kauffmann:1993aa,Cole:1994aa,  Somerville:1999aa}. Due to the limited achievable resolution, both approaches need to adopt parameterised models, and adjust the free parameters using observational  constraints. As these methods rely on empirical evidence, they are not fully `ab initio'. An alternative approach is taken by empirical models \citep{Mo:1996aa,  Wechsler:1998aa,Peacock:2000aa, Yang:2003aa,Zehavi:2004aa,Vale:2004aa,Conroy:2006aa,Moster:2010aa,Behroozi:2010aa,Guo:2010aa,Moster:2013aa,Moster:2018aa,Behroozi:2013aa,Behroozi:2019aa}, which use parameterised  relations between the properties of observed galaxies and those of simulated haloes. The parameters of these relations are constrained by requiring a number of statistical  observations to be reproduced. This approach has the advantage of accurately matching observations by construction, so that the evolution of observed galaxy properties can  be followed. We refer to the reviews by \citet{Naab:2016aa} for hydrodynamical simulations, \citet{Baugh:2006aa} and \citet{Somerville:2015aa} for SAMs, and  \citet{Wechsler:2018aa} for empirical models.

These very different approaches qualitatively agree that massive ETGs build up rapidly by gas dissipation and in-situ star formation at high redshift ($z >2$). At this phase, the galaxies are still star-forming and populate the `main sequence of star formation'. Towards lower redshifts the star formation ceases, the galaxies move to the `red sequence', and grow their stellar mass further by merging with ex-situ formed systems. Here the importance of ex-situ growth increases with mass and dominates above a certain mass scale. In models that implement the baryonic physics, like hydrodynamical simulations, the transition from star-forming to non-star-forming and the mass scale for accretion domination typically vary with the physical processes included but also with numerical resolution and the assumed sub-resolution model recipes, or simulation codes. The quenching of star formation and the existence of such a merger driven growth phase for massive ETGs, however, is clearly indicated by the observed evolution of galaxy stellar mass functions \citep[SMFs][]{Lilly:1995aa,Faber:2007aa} and evidence from size growth \citep{Trujillo:2006aa,vanDokkum:2008aa,vanderWel:2014aa}, outer stellar surface density profiles \citep{DSouza:2014aa}, present-day merger signatures \citep{Duc:2011aa}, or changes in the galaxy scaling relations for high mass systems \citep{Bernardi:2003aa}.

Here we present results on the formation and assembly histories of the ETG population of the empirical galaxy formation model \textsc{emerge}. This model follows the formation of individual dark matter haloes in cosmological simulations and links the growth rate of the halo to the star formation rate (SFR) of the galaxy at its centre. The parameterised conversion efficiency determines how efficiently baryons in the halo can be converted into stars. By construction the model successfully reproduces the evolution of the SMF, specific and cosmic star formation rates, the fractions of `quenched' galaxies, and mass-dependent galaxy correlation functions. We show that \textsc{emerge} predicts the observed evolution of the blue and red sequence with reasonable present-day archeological star formation histories for ETGs. Moreover, \textsc{emerge} reproduces the observed trend for present-day quenched galaxies to live in higher mass halos. Above a certain mass scale the growth of ETGs is dominated by merging and we compare this trend to current hydrodynamical simulations and a SAM. Finally we follow the main progenitors of present-day ETGs back in redshift and demonstrate their growing importance for the star-forming galaxy population up to redshift ($z \sim 8$).

This paper is organised as follows. In Section \ref{sec:method} we describe the methodology of \textsc{emerge} including the employed dark matter simulation and observational data. In Section \ref{sec:results} we present our main results: in section \ref{sec:shm} we present the stellar-to-halo mass (SHM) relation of active and passive galaxies and compare the model to the lensing constraints. In section \ref{sec:inexsitu} the predictions for the fraction of mass formed ex-situ are compared to other models. We determine the star formation and assembly histories of ETGs in section \ref{sec:sfh} and the build-up of the star formation bimodality in section \ref{sec:starformation}. In section \ref{sec:progenitors} we study the progenitors of ETG up to high redshift. Finally, we present our conclusions in section \ref{sec:conclusions}.

Throughout this work we assume a Planck $\Lambda{\rm CDM}$ cosmology with ($\Omega_\mathrm{m}$, $\Omega_\mathrm{\Lambda}$, $\Omega_\mathrm{b}$, $h$, $n_\mathrm{s}$, $\sigma_8$) = (0.308, 0.692, 0.0484, 0.6781, 0.9677, 0.8149). We employ a \citet{Chabrier:2003aa} initial mass function (IMF) and convert all stellar masses and SFRs to this IMF. All virial masses are computed according to the overdensity criterion by \citet{Bryan:1998aa}. To simplify the notation, we will use the capital $M$ to denote dark matter halo masses and the lower case $m$ to denote galaxy stellar masses. We use the terms ETG, passive galaxies, and quenched galaxies synonymously. Similarly, we use the terms LTG, active galaxies, and star-forming galaxies interchangeably.


\section{Method} \label{sec:method}

We follow the growth of galaxies in dark matter haloes with the empirical model \textsc{emerge} as presented in \citet{Moster:2018aa}. This code populates simulated halo merger trees with galaxies reproducing observed statistical galaxy properties such as stellar mass functions, star formation histories, and clustering. Consequently, the two main pillars of the model are cosmological $N$-body simulations from which dark matter haloes and merger trees are extracted, and observed statistical galaxy data. In this section, we give a short summary of those components and how \textsc{emerge} relates them. More details can be found in the model paper \citep{Moster:2018aa}.


\subsection{Simulation} \label{sec:simulation}

The model \textsc{emerge} traces dark matter haloes through cosmic time and assigns a SFR to the galaxy at its centre depending on its mass and growth rate. We have extracted halo merger trees from a cosmological $N$-body simulation with $200\Mpc$ side length. The adopted cosmological parameters are consistent with the latest results by the \citet{Planck-Collaboration:2018aa}: $\Omegam=0.308$, $\OmegaL=0.692$, $\Omegab=0.0484$, $H_0=67.81\kmsMpc$, $n_\mathrm{s}=0.9677$, and $\sigma_8=0.8149$. The initial conditions for the simulation were generated with the {\sc Music} code \citep{Hahn:2011aa} and a power spectrum computed with the \texttt{CAMB} code \citep*{Lewis:2000aa}. The simulation contains $1024^3$ collisionless particles with a mass of $2.92\times10^8\Msun$ and was run with periodic boundary conditions from redshift $z=63$ to 0 using the TreePM code {\sc Gadget3} \citep{Springel:2005aa} with a gravitational softening of $3.3\kpc$. We saved 94 snapshots which are equally spaced in scale factor ($\Delta a=0.01$). Using the halo finder {\sc Rockstar} \citep*{Behroozi:2013ac}, we identified the dark matter haloes and subhaloes in each snapshot. Halo masses were derived using the criterion by \citet{Bryan:1998aa}. With a minimal particle number of 100 for each halo the minimally resolved halo mass is $\log_{10}(M_\mathrm{min}/\Msun)=10.5$. Merger trees were generated with the {\sc ConsistentTrees} code \citep{Behroozi:2013ad}. The term `main halo' is used to refer to distinct haloes that are not located within a larger halo, while all other haloes are called `subhalos'. We further assume that a `central galaxy' is located at the centre of a main halo, and a `satellite galaxy' within each subhalo.


\subsection{Observations} \label{sec:observations}

Given the properties of the halo as extracted from the simulation, the model assigns properties to its galaxy such that a range of observed statistical galaxy properties are reproduced. The observational constraints used in this work are SMFs, cosmic SFR densities, specific SFRs, quenched galaxy fractions, and projected galaxy correlation functions. All units are converted into physical units with $h=0.6781$ and a \citet{Chabrier:2003aa} IMF. Systematic errors are taken into account by calculating the variance between different observations and adding the result quadratically to each data point. For the SMFs this error is $\sigma=0.15 {\rm~dex}$ at $z\le0.5$ and $\sigma=0.3 {\rm~dex}$ beyond. The error found for the cosmic SFR density is $\sigma=0.1 {\rm~dex}$, while for the specific SFRs we find $\sigma=0.15 {\rm~dex}$, independent of stellar mass and redshift. For the systematic error of the quenched fractions we use $\sigma=10$ per cent, while for the clustering we use $\sigma=0.15 {\rm~dex}$. We have used all the data that was used in \citep{Moster:2018aa}, and refer to their tables 1 to 5 for more information. In addition we have have used the SMFs by \citet{Wright:2018aa}, \citet{Beare:2019aa} and \citet{Bhatawdekar:2019aa}, the cosmic SFR densities by \citet{Katsianis:2017aa} and \citet{Liu:2018aa}, and the specific SFRs by \citet{Davidzon:2018aa}. We account for intrinsic scatter in the mass estimates relative to the true mass arising from limited photometric information by drawing stellar masses from a lognormal distribution with a mean value given by the model mass and a scatter of $\sigma(z)=0.08+0.06z$, capped at a maximum value of $0.32 {\rm~dex}$ at $z=4$, when comparing to observed stellar masses.


\subsection{Galaxy formation model} \label{sec:model}

The simulated dark matter haloes are populated with galaxies by the empirical model \textsc{emerge}, such that the observed data is reproduced. To this end, the SFR of the galaxy at the centre of a dark matter halo is set to the product of the halo's baryonic growth rate and the instantaneous baryon conversion efficiency $\epsilon$:
\begin{equation} \label{eqn:sfrcen}
\frac{{\rm d}m_*}{{\rm d}t} (M,z) = \frac{{\rm d}m_\mathrm{b}}{{\rm d}t} \cdot \, \epsilon(M,z) = f_\mathrm{b} \frac{{\rm d}M}{{\rm d}t} \cdot \, \epsilon(M,z)\; ,
\end{equation}
where $f_\mathrm{b} = \Omegab/\Omegam$ is the universal baryonic fraction. In this way, the baryonic growth rate signifies how much material becomes available for galaxy formation, while the conversion efficiency determines how effectively it is converted into stars. The halo growth rates are computed from the merger trees and averaged over one dynamical time. For negative halo growth rates we set the baryonic growth rate to zero, and we do not correct for pseudo-evolution.

The conversion efficiency $\epsilon$ describes how efficiently the gas that falls into the halo is converted into stars, and thus combines the effects of cooling, star formation, and various feedback processes. Following an Occam's razor approach to find the simplest model that is able to reproduce the observed data, we have tested several parametrizations  for the efficiency, which we assessed with various model selection criteria, such as the Bayesian Evidence. Consequently, we parametrise the efficiency as a double power law with normalisation $\epsilon_\mathrm{N}$, characteristic halo mass $M_1$, and slopes $\beta$ and $\gamma$:
\begin{equation} \label{eqn:epsilon}
\epsilon(M,z) = 2 \;\epsilon_\mathrm{N}(z) \left[ \left(\frac{M}{M_1(z)}\right)^{-\beta(z)} + \left(\frac{M}{M_1(z)}\right)^{\gamma(z)}\right]^{-1} \; ,
\end{equation}
where $\epsilon_\mathrm{N}$, $M_1$, and $\beta$ depend linearly on the scale factor $a$, and $\gamma$ is a constant:
\begin{align}
\log_{10} M_1(z)& = M_0 + M_\mathrm{z}(1-a) = M_0 + M_\mathrm{z}\frac{z}{z+1} \; ,\\
\epsilon_\mathrm{N}(z)& = \epsilon_0 + \epsilon_\mathrm{z}(1-a) = \epsilon_0 + \epsilon_\mathrm{z}\frac{z}{z+1} \; ,\\
\beta(z)& = \beta_0 + \beta_\mathrm{z}(1-a) = \beta_0 + \beta_\mathrm{z}\frac{z}{z+1} \; ,\\
\gamma(z)& = \gamma_0 \; .
\end{align}
While we do not impose any a-priori restrictions on the values of the slopes, we find the values of $\beta$ and $\gamma$ to be positive at all cosmic times, i.e. at low/high masses the efficiency increases/decreases with increasing halo mass.

Once the growth rate $\dot M$ and the conversion efficiency $\epsilon(M,z)$ have been computed for a halo at a specific time, \textsc{emerge} can compute the SFR of the central galaxy $\dot m_*(M,\dot M,z)$ using equation ($\ref{eqn:sfrcen}$). Integrating this rate over cosmic time, while taking into account the fraction of mass that is being lost as a consequence of dying stars, yields the stellar mass formed in-situ. In addition, the stellar mass of central galaxies grows by the accretion of satellite galaxies, i.e. by ex-situ mass added during mergers. In our model, galaxies in haloes with the same mass and growth rate at a given redshift have the same SFR, but the stellar mass is different depending on the full formation history of the haloes. The star formation history of each galaxy is then a consequence of the specific path a halo has taken through the halo mass-redshift plane. Typically, at early cosmic times the SFR of a galaxy will be low, as both the growth rate of its halo and the conversion efficiency are low. When the halo mass approaches the characteristic mass $M_1$, the conversion efficiency is maximal, resulting in the peak of the star formation history. As the halo becomes more massive, the conversion efficiency decreases again, resulting in quenched central galaxies.

For growing dark matter haloes, the evolution of the stellar content is determined by the conversion efficiency. Once a halo starts to lose mass, which roughly corresponds to the time its galaxy becomes a satellite, we consider three physical processes that regulate galactic growth: satellite quenching, stripping and merging. For each process the simplest model that is able to reproduce the observational data has been chosen, though the complexity of the models can be increased if new observational data require it.

Once a halo starts to be accreted by a larger halo, its mass begins to decline as a result of tidal stripping. Consequently the infall of gas onto its galaxy stops, such that the cold gas reservoir gets used up by star formation, quenching the galaxy. We take this process into account by employing the `delayed-then-rapid' quenching model, i.e. once the halo stops growing, the SFR of its galaxy is kept constant for a time $\tau$ and then set to zero. This quenching time is parameterised with respect to the halo's dynamical time $t_\mathrm{dyn}=R_\mathrm{vir}/v_\mathrm{vir}$ and is longer for low mass satellites:
\begin{equation} \label{eqn:satquenching}
\tau = t_\mathrm{dyn} \cdot \tau_0 \cdot \max \left[ \left(\frac{m_*}{10^{10}\Msun}\right)^{-\tau_\mathrm{s}},1\right]  \; .
\end{equation}
The parameters $\tau_0$ and $\tau_\mathrm{s}$ are mostly constrained by the observed fraction of quenched galaxies, as shorter quenching times lead to higher quenched fractions. If the halo mass surpasses its previous peak, the halo growth rate and the instantaneous conversion efficiency again determine the SFR (eqn. $\ref{eqn:sfrcen}$).

If a subhalo has lost enough of its mass through tidal stripping, its gravitational potential is no longer able to protect the stars in its centre from stripping, such the galaxy is destroyed. This process is implemented by moving its stars to the stellar halo (ICM) around the central galaxy once the mass of its subhalo has dropped below a fraction $f_\mathrm{s}$ of the peak mass:
\begin{equation} \label{eqn:satstripping}
M < f_\mathrm{s} \cdot M_\mathrm{peak}
\end{equation}
The stripping parameter $f_\mathrm{s}$ is mostly constrained by small-scale clustering, as faster stripping leads to fewer satellites on small scales, and consequently to lower small-scale clustering. Once a halo becomes a subhalo, we move all the mass in its ICM to the ICM of the host halo.

\begin{table}
 \caption{Fitting results from MCMC}
 \label{tab:bestfit}
 \begin{tabular}{@{}lccc@{}}
  \hline
  Parameter & Best-fit & Upper 1$\sigma$ & Lower 1$\sigma$\\
  \hline
  $M_0$ & 11.339 & +0.005 & -0.008\\
  $M_z$ & ~0.692 & +0.010 & -0.009\\
  $\epsilon_0$ & ~0.005 & +0.001 & -0.001\\
  $\epsilon_z$ & ~0.689 & +0.003 & -0.003\\
  $\beta_0$ & ~3.344 & +0.084 & -0.101\\
  $\beta_z$ & -2.079 & +0.127 & -0.134\\
  $\gamma_0$ & ~0.966 & +0.002 & -0.003\\
  \hline
  $f_\mathrm{esc}$ & ~0.388 & +0.002 & -0.002\\
  $f_\mathrm{s}$ & ~0.122 & +0.001 & -0.001\\
  $\tau_0$ & ~4.282 & +0.015 & -0.020\\
  $\tau_\mathrm{s}$ & ~0.363 & +0.014 & -0.014\\
  \hline
  \end{tabular}
 \medskip\\
  \textbf{Notes:} All masses are in units of \Msun.
\end{table}

Once a subhalo has lost all of its orbital energy due to dynamical friction, the satellite galaxy will merge with the central galaxy. At this point we let a fraction of satellite stars $f_\mathrm{esc}$ escape to the halo as diffuse stellar material, resulting in a  remnant mass of
\begin{equation} \label{eqn:satmerging}
m_\mathrm{rem} = m_\mathrm{cen} + m_\mathrm{sat} \cdot (1-f_\mathrm{esc}) \; ,
\end{equation}
and an ICM mass of $m_\mathrm{ICM,new}=m_\mathrm{ICM,old}+f_\mathrm{esc} m_\mathrm{sat}$. The escape fraction $f_\mathrm{esc}$ is mostly constrained by the evolution of the massive end of the SMF at low redshift, as a high escape fraction will lead to a slow growth of massive galaxies.

When the mass of a subhalo has been stripped below the resolution limit of the simulation, it can no longer be identified and its galaxy becomes an `orphan'. We keep the orphans until their dynamical friction time $t_\mathrm{df}$ has elapsed \citep[employing][]{Boylan-Kolchin:2008aa} and then let them merge with their central galaxies. If a host halo of an orphan merges with a larger halo, we recompute the dynamical friction time with respect to the new central galaxy, and let the orphan merge with it once this new time has elapsed.  

To constrain the model parameters we construct mock observations and calculate a model value for each individual observed data point, so that we can directly compare both. To calculate the quenched fractions, we define a galaxy to be quenched if its sSFR is below a redshift dependent threshold given by $\Psi<0.3t_\mathrm{H}^{-1}$, where $t_\mathrm{H}$ is the Hubble time at that redshift. We calculate the real space galaxy two-point correlation functions $\xi(r)$ using kd-trees following \citet{Moore:2001aa} and \citet{Dolence:2008aa}, and derive the projected correlation functions $w_\mathrm{p}(r_\mathrm{p})$ at the same projected radii as the observations. We compare the observed data points $\vec \omega$ to the modelled values $\vec \mu (\vec \theta)$ for a set of parameters $\vec \theta$ to get the difference $\vec \Delta = \vec \omega - \vec \mu (\vec \theta)$. With the covariance matrix of the observed data $C$, we then compute $\chi^2 = \vec\Delta^T \, C^{-1} \, \vec\Delta$, and assign a likelihood to the model $\mathcal{L} = \exp(-\chi^2/2)$. Parameter space is then explored with two different sampling techniques \citep{Elson:2007aa,Goodman:2010aa} to find the most likely parameters and their credibility intervals, which are presented in Table \ref{tab:bestfit}.

\begin{figure}
    \includegraphics[width=\halfwidth]{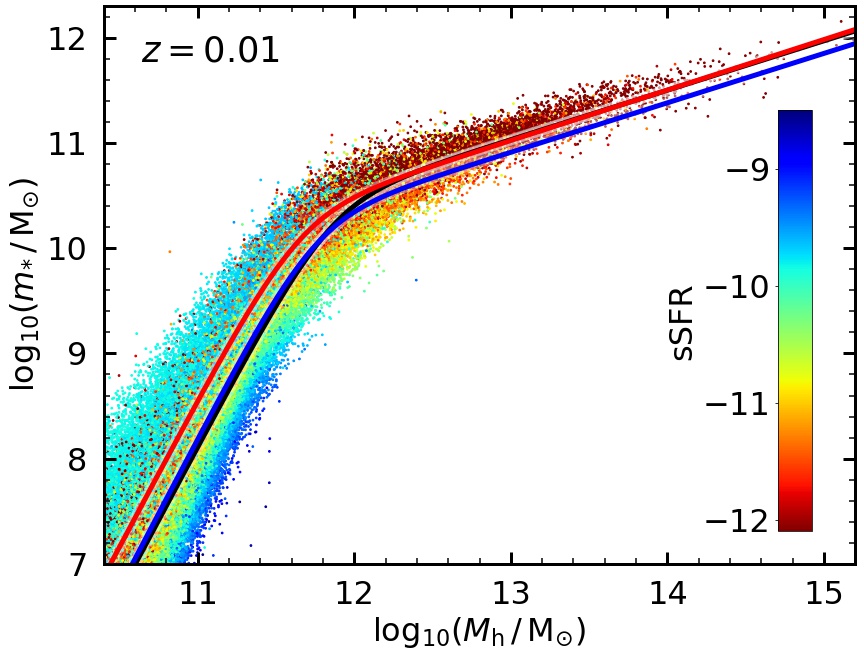}
    \caption{
        Relation between stellar mass $m_*$ and peak halo mass $M_\mathrm{h}$ for individual central galaxies at $z=0.01$. The colour of each point corresponds to the specific SFR ($\mathrm{yr}^{-1}$) of the galaxy as indicated by the colour bar. The solid black line shows the median stellar mass at fixed halo mass, while the median stellar mass for active and passive central galaxies are given by the blue and red line, respectively. At fixed halo mass, passive galaxies are more massive than active galaxies.
    }
    \label{fig:shm}
\end{figure}

\begin{figure}
    \includegraphics[width=\halfwidth]{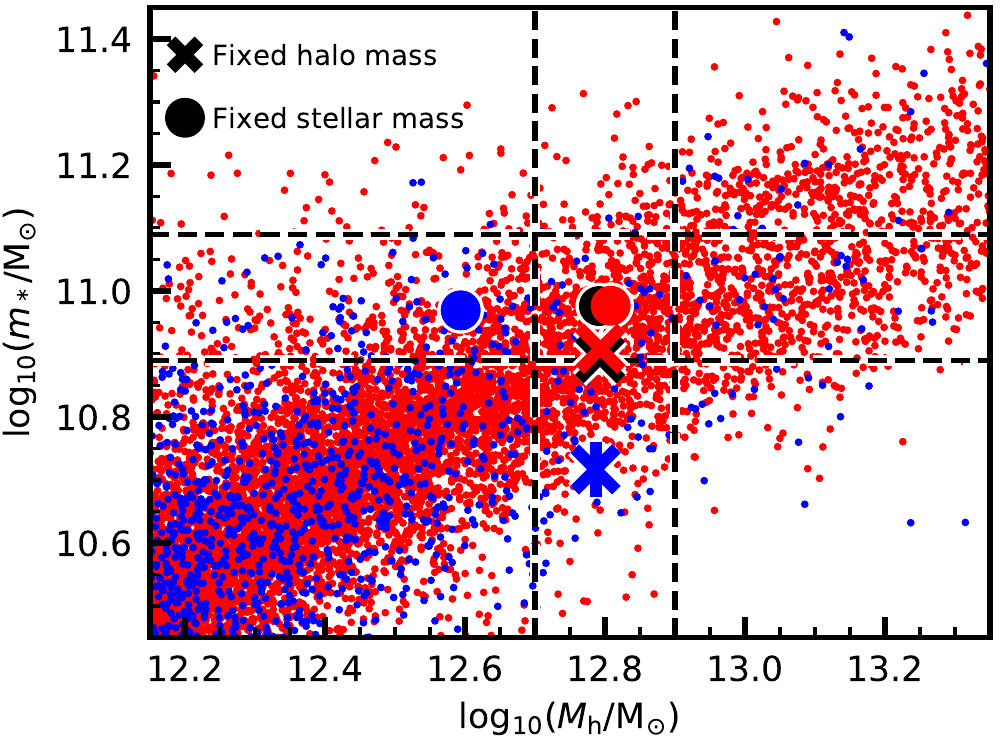}
    \caption{
    The relation between stellar and halo mass for central galaxies computed in bins of fixed halo and stellar mass. Blue and red dots represent individual active and passive galaxies, respectively. The vertical dashed lines indicate the chosen halo mass bin, while the vertical dashed lines indicate the chosen stellar mass bin. The crosses show the average stellar and halo mass within the halo mass bin, while the filled circles give the average stellar and halo mass within the stellar mass bin. Black, blue, and red symbols represent all, active, and passive central galaxies, respectively.
    }
    \label{fig:mhmsbin}
\end{figure}

To efficiently explore parameter space even for large dark matter simulations, \textsc{emerge} has been adapted to run on massively parallel computers with distributed memory. First the halo merger trees are loaded and distributed among a set of $N_\mathrm{C}$ computing cores (processors) using the Message Passage Interface (MPI) protocol. The trees in this first `universe' are copied to more sets of $N_\mathrm{C}$ cores with MPI, to create a total of $N_\mathrm{U}$ parallel universes on total number of $N_\mathrm{U} \cdot N_\mathrm{C}$ cores. Parameter space is then explored with $N_\mathrm{W}$ walkers, such that at any time $N_\mathrm{U}$ walkers are processed in parallel, requiring $N_\mathrm{W} / N_\mathrm{U}$ steps for all walkers. Each core first computes the galaxy properties for all haloes in its trees, which can be further parallelised with OpenMP if the cores support hyperthreading. All cores in one universe combine their information to calculate the model statistics and the resulting model likelihood. Combining all these different parallelisation techniques, \textsc{emerge} is able to achieve a near-perfect scaling. For our $200\Mpc$ box and Xeon Gold 2GHz CPUs, around 30 universes can be computed per core hour. The source code and documentation of \textsc{emerge} are available on GitHub\footnote{github.com/bmoster/emerge}.


\section{Results} \label{sec:results}

The empirical model \textsc{emerge} was designed to follow the assembly of galaxies in dark matter haloes while being in good agreement with global statistics, i.e. SMFs, cosmic SFR densities,  specific SFRs, quenched fractions, and galaxy clustering. We will use this model to investigate the different formation mechanisms of active and passive galaxies, and to study the progenitors of ETGs.


\subsection{The SHM relation for active and passive galaxies} \label{sec:shm}

The SFR of a galaxy is given by the product of the growth rate of its halo and the conversion efficiency, which in turn depends on the halo's mass and redshift. Therefore, the stellar mass of a galaxy is a direct consequence of the individual formation history of its dark matter halo. In M13 we have shown that independent of halo mass, passive galaxies embedded in haloes of a specific mass have formed more stellar mass than their active counterparts. At fixed halo mass, the conversion efficiency is the same in all haloes, so that the SFR of a galaxy only depends on the halo's growth rate. Haloes with a low growth rate thus host quenched galaxies, but they must have had a high growth rate at earlier times to reach the same final halo mass, which means that they hosted galaxies with a high (specific) SFR, i.e. active galaxies. Conversely, haloes with high growth rates at late times host active galaxies, but must have had low growth rates (hosting passive galaxies) at earlier times to reach the same final halo masses. However, as the conversion efficiency evolves with redshift, this does not imply that also the stellar mass reaches the same value at late times. Since the conversion efficiency is higher at high redshift, haloes with high growth rates at early times host galaxies with very high specific SFRs, such that the resulting final stellar mass is higher compared to galaxies in haloes with low growth rates at high redshift. While those haloes have higher growth rates (and host active galaxies) at late times, the lower conversion efficiency results in SFRs that cannot compensate the lower SFRs at high redshift, such that the final stellar masses are lower. We show the resulting SHM relation for each galaxy and the mean stellar mass at fixed halo mass for active, passive, and all central galaxies in Figure \ref{fig:shm}.

\begin{figure}
    \includegraphics[width=\halfwidth]{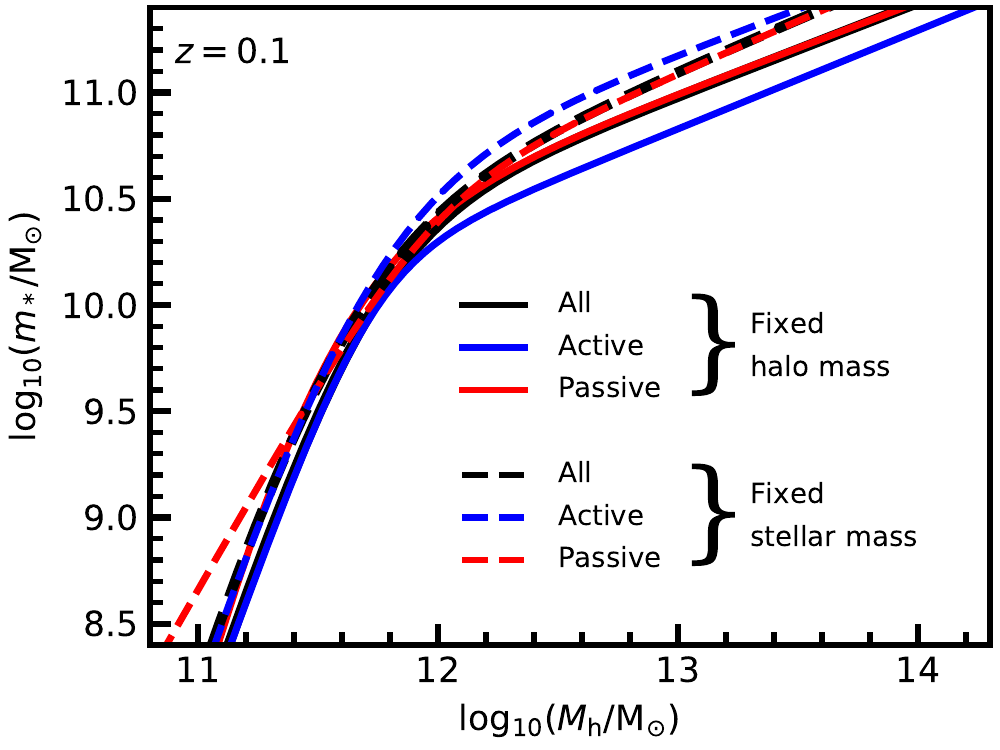}
    \caption{
    Comparison of the central galaxy stellar-to-halo mass relation for fixed halo and stellar mass. The solid lines show the relation at fixed halo mass, while the dashed lines give the relation at fixed stellar mass. Black, blue, and red lines represent all, active, and passive galaxies, respectively.
    }
    \label{fig:mhmscomp}
\end{figure}

\begin{figure}
    \includegraphics[width=\halfwidth]{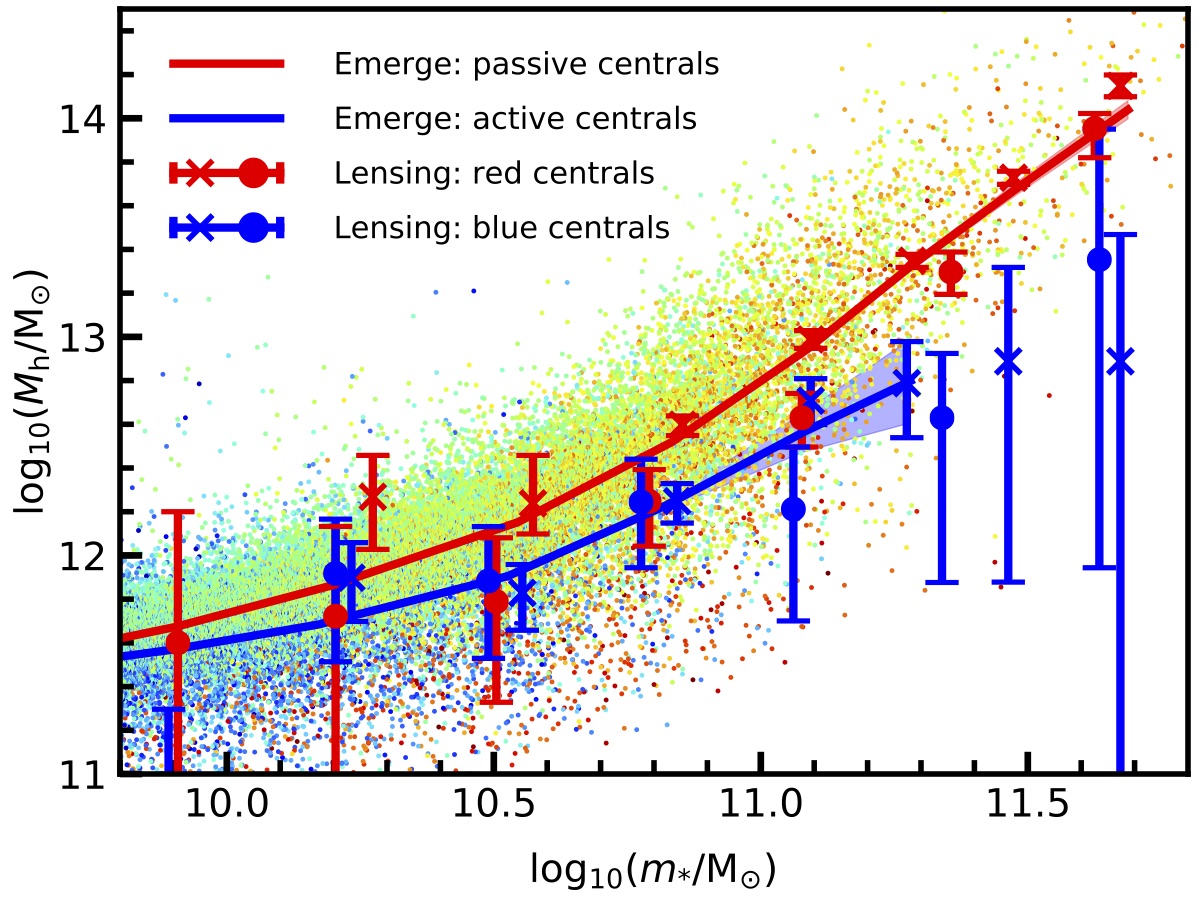}
    \caption{
      Comparison of the stellar-to-halo mass relation averaged at fixed stellar mass for active and passive galaxies between \textsc{emerge} (solid lines), and the results of weak lensing (symbols with error bars). The filled circles are taken from \citet{Mandelbaum:2006aa} and the crosses from \citet{Mandelbaum:2016aa}. The dots in the background show the relation for individual galaxies computed by \textsc{emerge} using the same colour-coding for the specific SFR as in Figure \ref{fig:shm}. The model and the data are in very good agreement and show that passive galaxies tend to live in more massive haloes than active galaxies.
    }
    \label{fig:msmh}
\end{figure}

Our model \textsc{emerge} thus predicts that at fixed halo mass, passive galaxies have a higher stellar mass than active galaxies. This result seems to be in disagreement with the findings of weak lensing studies \citep[e.g.][]{Mandelbaum:2006aa,Mandelbaum:2016aa}, which show that at fixed stellar mass, passive galaxies tend to live in more massive haloes than active galaxies. However, simply inverting the average SHM relation at fixed halo mass does not result in the correct SHM relation at fixed halo mass because of the scatter. We demonstrate this in Figure \ref{fig:mhmsbin}, where we have colour-coded all active galaxies in blue and all passive galaxies in red. The vertical dashed lines indicate a halo mass bin between $\log M_\mathrm{h}/\Msun=12.7$ and $12.9$, while the black, red, and blue crosses show the average stellar mass of all, active, and passive galaxies in this halo mass bin, respectively. As we noted above, passive galaxies have a higher stellar mass than active galaxies ($\log m_*/\Msun=10.9$ vs. $10.7$). In this rather high halo mass bin, most galaxies are quenched, so that the average stellar mass of all galaxies is close to the average mass of passive galaxies.

The horizontal dashed lines in Figure \ref{fig:shm} indicate a stellar mass bin with the same width as the halo mass bin (0.2 dex). The bin edges $\log m_*/\Msun=10.9$ and $11.1$ have been chosen, such that the average halo mass for all galaxies is the same as the average halo mass of the halo mass bin ($\log M_\mathrm{h}/\Msun=12.8$). Here, we notice the opposite trend, i.e. active galaxies are located in lower mass haloes than passive galaxies ($\log M_\mathrm{h}/\Msun=12.6$ vs. $12.8$). This is a consequence of two effects: first, the fraction of active galaxies is higher at lower halo (and stellar) masses, and second, the scatter in stellar mass at fixed halo mass is also higher at lower halo mass. Consequently, there are more active galaxies in low-mass haloes that are scattered to higher stellar masses compared to active galaxies in massive haloes that are scattered to lower masses. We further note, that for the same halo mass the average stellar mass is higher at fixed halo mass than at fixed stellar mass. This also results from a combination of higher scatter at lower halo masses and a larger number of low-mass haloes, such that more galaxies are scattered up from lower halo masses than scattered down from higher halo masses. The same stellar mass is thus reached at lower halo mass, which leads to a higher SHM ratio.

We further investigate difference in the SHM relation at fixed halo mass and fixed stellar mass in Figure \ref{fig:mhmscomp}. For all masses, the SHM ratio is higher when computed at fixed stellar mass compared to the ratio obtained at fixed halo mass. At the massive end, the SHM ratio at fixed halo mass is lower for active galaxies compared to passive galaxies, while the SHM ratio at fixed stellar mass is higher for active galaxies than for passive galaxies. At intermediate masses, the SHM relation is very similar for active and passive galaxies, both computed at fixed stellar and fixed halo mass. At the low-mass end we find that the SHM ratio at fixed halo mass is still higher for passive galaxies, but we also find that the SHM ratio at fixed stellar mass is higher for passive galaxies, in contrast to the SHM ratio at the massive end. This means that \textsc{emerge} predicts that at the massive end, passive galaxies of given stellar mass are located in more massive haloes than active galaxies, while at the low-mass end, passive galaxies are hosted by less massive haloes than active galaxies. The trend for fixed stellar mass reverses at $\log M_\mathrm{h}/\Msun \approx 11.5$, and $\log m_*/\Msun \approx 9.5$.

Finally, we compare the predictions of \textsc{emerge} directly to the observationally measured SHM relation at fixed stellar mass. To this end, we compute the average halo mass for active and passive galaxies in the same stellar mass bins as \citet{Mandelbaum:2016aa}. For each bin, we use the model at the same redshift as the median redshift in the observations. We convert all halo masses to our definition. The resulting SHM relations for active and passive galaxies and the data by \citet{Mandelbaum:2016aa} and \citet{Mandelbaum:2006aa} are presented in Figure \ref{fig:msmh}. The predictions of \textsc{emerge} are in very good agreement with the lensing constraints. As noted before, the SHM ratio for galaxies with $\log m_*/\Msun > 9.5$ is lower for passive galaxies, i.e. at the same stellar mass, they live in more massive haloes than their active counterparts. The model further predicts that this trend inverts for lower masses. Overall, we take this agreement with the data as a further confirmation of the model's objective, i.e. to link galaxies to dark matter haloes such that observational data is reproduced.


\subsection{The growth of massive galaxies through accretion} \label{sec:inexsitu}

\begin{figure}
    \includegraphics[width=\halfwidth]{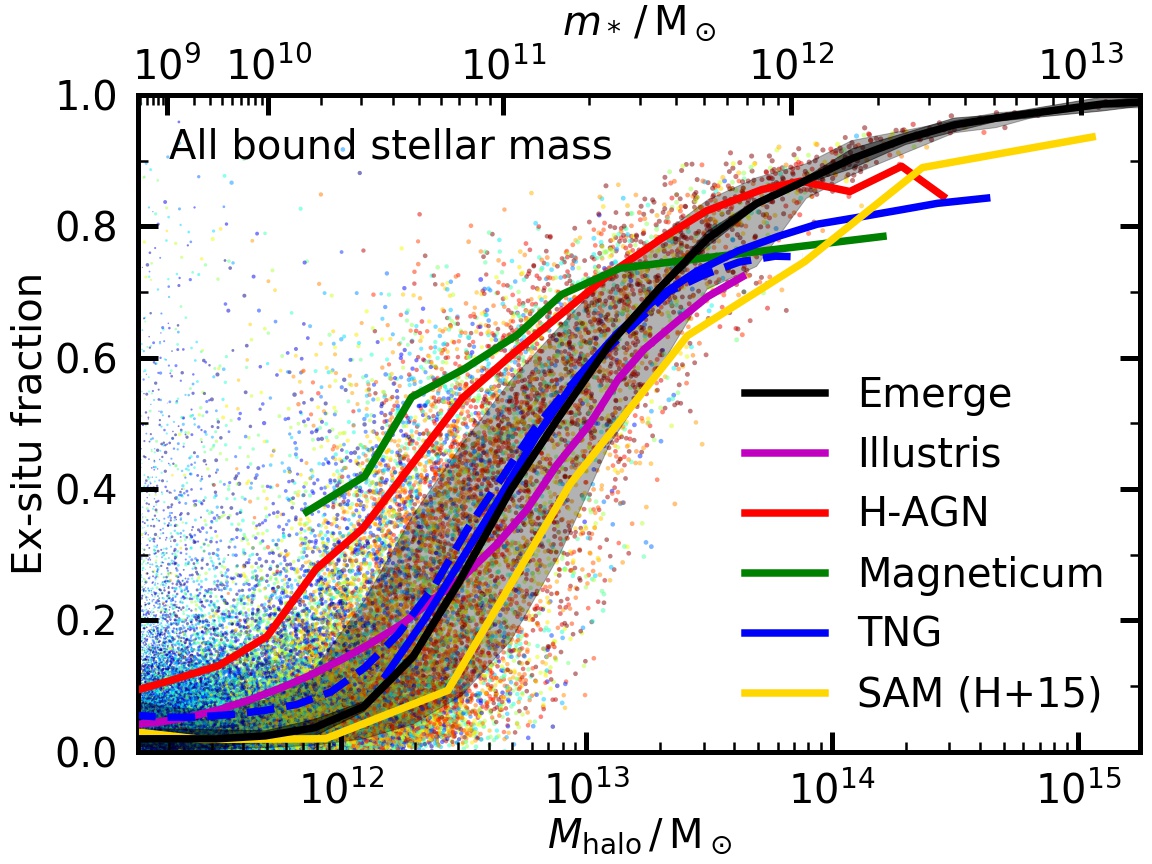}
    \caption{
      The fraction of stellar mass formed ex-situ as a function of halo mass for all bound stellar mass within the dark matter halo (central galaxy and ICM) at $z=0$. The axis on the top indicates the bound stellar mass that corresponds to the bottom axis. The dots in the background show the ex-situ mass fraction for individual galaxies computed by \textsc{emerge} using the same colour-coding for the specific SFR as in Figure \ref{fig:shm}. The black line shows the average ex-situ fraction for \textsc{emerge}. The light grey area indicates the $1\sigma$ scatter in the relation. The dark grey area represents the uncertainty in the average relation resulting from observational uncertainties and has been computed from all MCMC walkers. In comparison, we show the results of hydrodynamical simulations and a SAM (coloured lines).
    }
    \label{fig:fexAll}
\end{figure}

\begin{figure*}
    \includegraphics[width=\fullwidth]{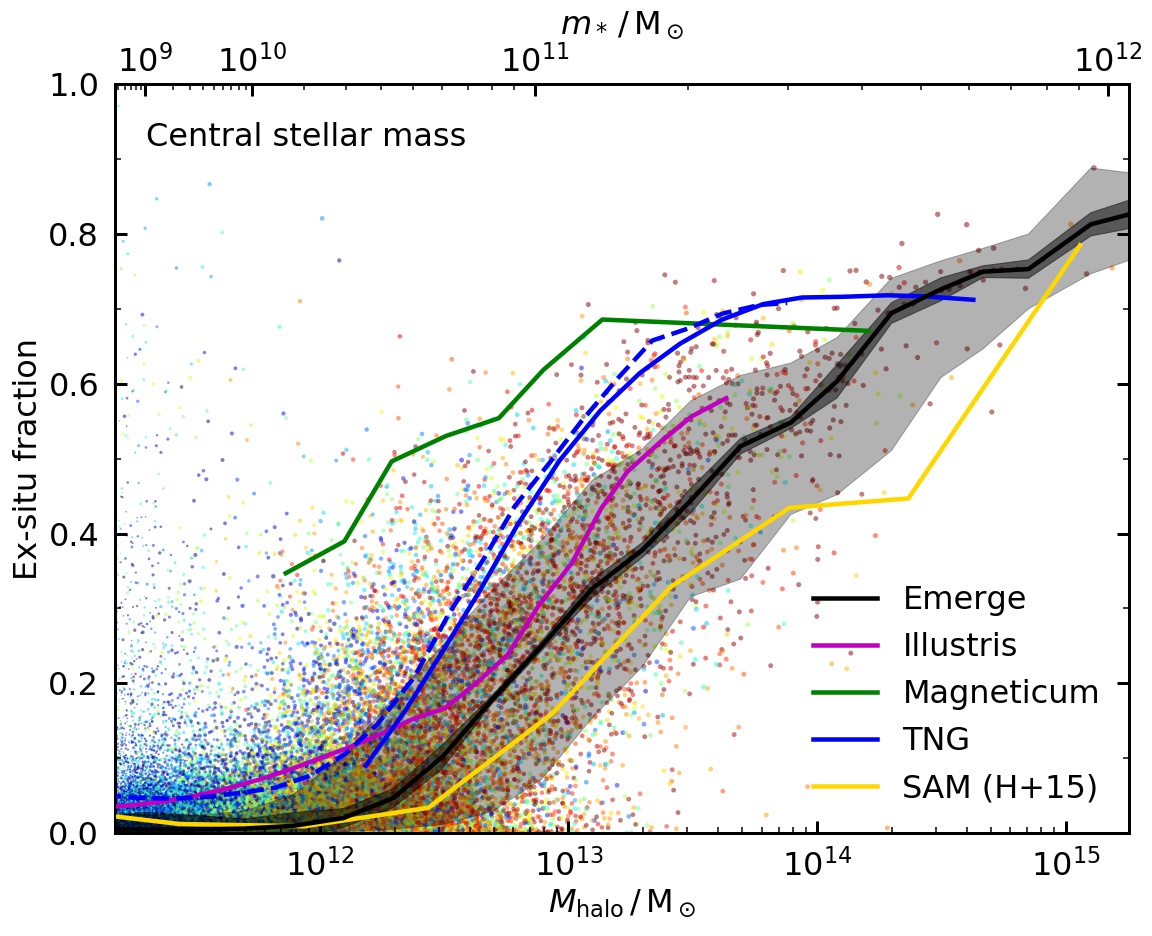}
    \caption{
       Similar to Figure \ref{fig:fexAll}, but for the stellar mass of the central galaxy excluding the ICM. The axis on the top indicates the central galaxy stellar mass that corresponds to the bottom axis. The dots in the background show the ex-situ mass fraction for individual galaxies computed by \textsc{emerge}, and the black line shows the average ex-situ fraction. In comparison, we show the results of hydrodynamical simulations and a SAM (coloured lines).
    }
    \label{fig:fexCen}
\end{figure*}

In a hierarchical structure formation scenario, dark matter haloes grow by continuously accreting smaller haloes. In this picture, galaxies do not only gain mass by converting their accreted gas into stars (in-situ growth), but also by accreting other galaxies (ex-situ growth). To better understand how massive galaxies assemble their mass, we compute the ex-situ mass fraction $f_\mathrm{ex}$ for each individual galaxy at $z=0$, as the ratio between the remaining stellar mass that has been accreted and the remaining total stellar mass. We show the results for all bound stellar mass, i.e. mass in the central galaxy plus the mass in the ICM, in Figure \ref{fig:fexAll}, and the results for the stellar mass in the central galaxy in Figure \ref{fig:fexCen}. In both Figures, the lower abscissa indicates the peak halo mass for each galaxy, while the upper abscissa indicates the stellar mass (total bound for Figure \ref{fig:fexAll} and central for Figure \ref{fig:fexCen}) with an average halo mass that corresponds to the lower abscissa. While this is not directly equivalent to the stellar mass of each individual galaxy because of scatter, the upper abscissa still yields a good proxy for how the ex-situ mass fraction scales with stellar mass. The dots represent individual systems as computed by \textsc{emerge}, while the black solid line gives the average $f_\mathrm{ex}$ for a fixed halo mass. The wide light grey area indicates the $1\sigma$ scatter in this relation, and the narrow dark grey area gives the uncertainty of the average resulting from observational uncertainties. The latter has been derived by computing the average ex-situ mass fraction for all final MCMC walkers, and then calculating the standard deviation.

\begin{figure*}
    \includegraphics[width=\halfwidth]{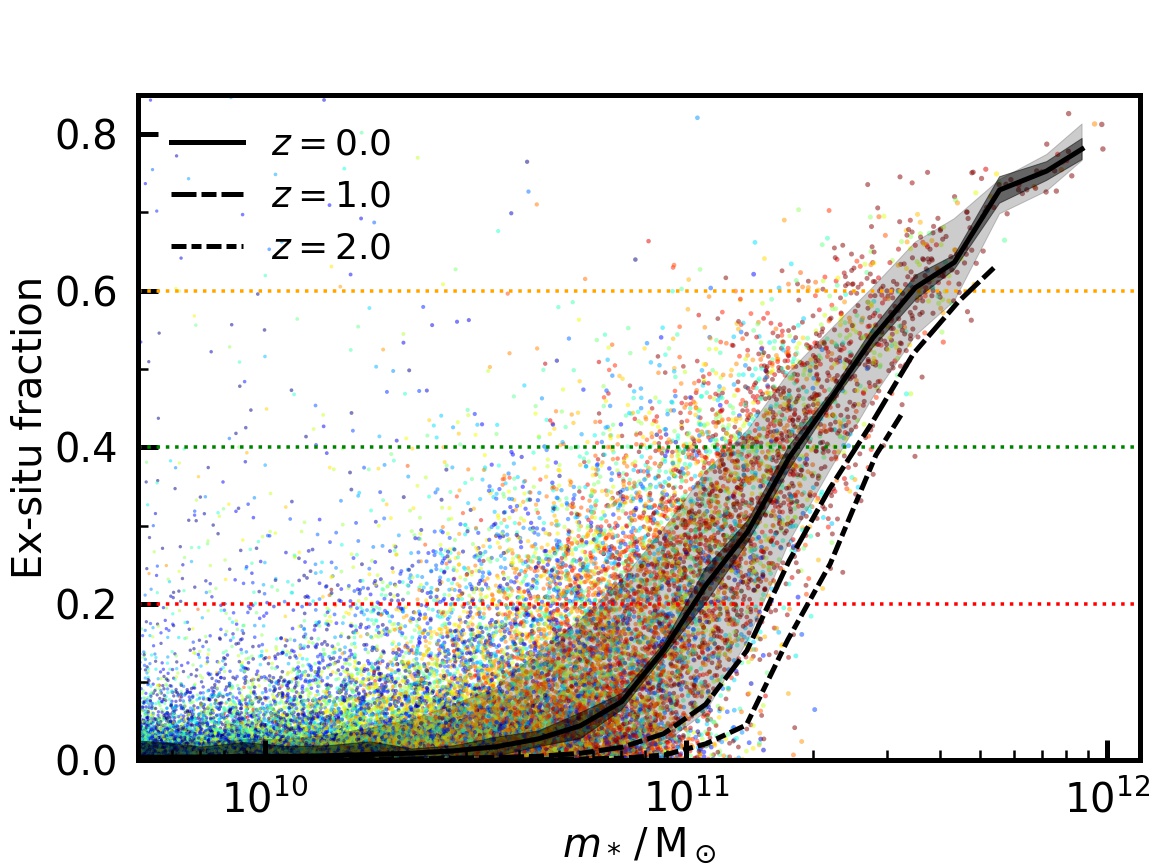}
    \hspace{5mm}
    \includegraphics[width=\halfwidth]{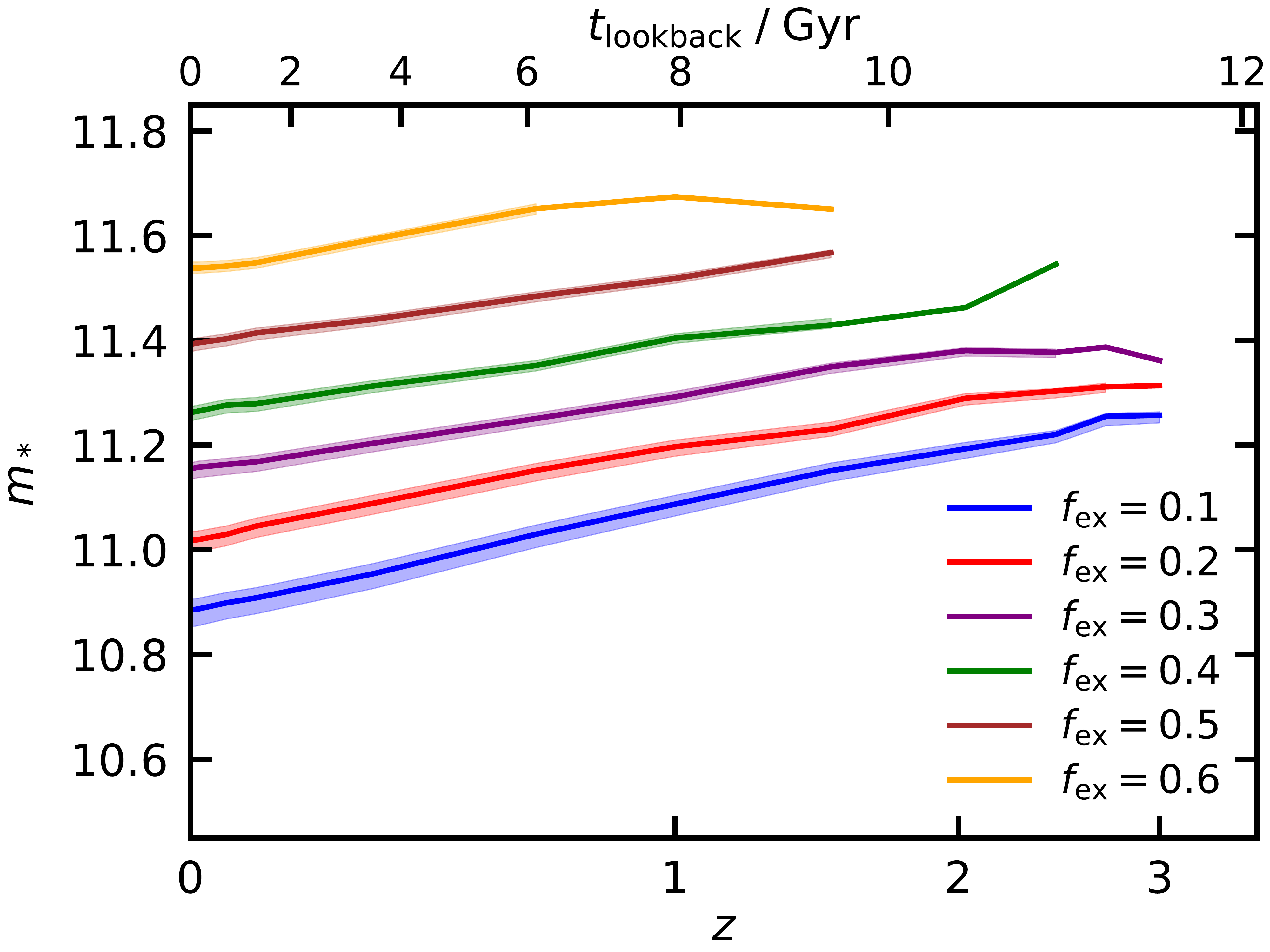}
    \caption{The stellar mass at which growth through accretion becomes significant.
    \textit{Left-hand panel:} The ex-situ mass fraction of central galaxies as a function of stellar mass. The dots show the ex-situ mass fraction for individual galaxies at $z=0$, and the black lines show the average ex-situ fraction at different redshifts. The horizontal dotted lines indicate three specific ex-situ fractions.   
    \textit{Right-hand panel:} Stellar mass of galaxies that have accreted a given fraction of their total stellar mass as a function of redshift. The lines with different colours show the results of \textsc{emerge} for different fixed ex-situ fractions, e.g. at $z=0$ central galaxies with $\log (m_*/\Msun)=11$ have accreted 20 per cent of their total stellar mass, while at $z=2$ the stellar mass that corresponds to an ex-situ fraction of 20 per cent is $\log (m_*/\Msun)=11.3$.
    }
    \label{fig:macc}
\end{figure*}

The average ex-situ mass fraction for all bound mass is very low in low-mass haloes with $\log(M_\mathrm{h}/\Msun)<12$, although there are individual galaxies with $f_\mathrm{ex}$ as high as 90 per cent. Starting at $\log(M_\mathrm{h}/\Msun)=12$ where $f_\mathrm{ex}$ is 5 per cent, the ex-situ mass fraction increases quickly and reaches 57 per cent at $\log(M_\mathrm{h}/\Msun)=13$ and 89 per cent at $\log(M_\mathrm{h}/\Msun)=14$. Beyond this mass, $f_\mathrm{ex}$ increases more slowly and approaches 100 per cent. We contrast this with the results of hydrodynamical simulations including Illustris \citep[magenta line;][]{Vogelsberger:2014aa}, Horizon-AGN \citep[red line][]{Dubois:2014aa}, Magneticum \citep[green line;][]{Hirschmann:2014aa}, and Illustris TNG \citep[solid blue line: TNG300, dashed blue line: TNG100;][]{Pillepich:2018aa}, and we convert all halo masses to our definition. We also include the semi-analytic model by \citet{Henriques:2019aa} in our comparison (yellow line). For low-mass haloes, we find a very good agreement with the results of Illustris and TNG, although they predict slightly higher ex-situ fraction than \textsc{emerge}. For more massive haloes, Illustris has slightly lower values for $f_\mathrm{ex}$. Similarly, TNG follows the empirical prediction very closely below $\log(M_\mathrm{h}/\Msun)\le13.5$, but above this halo mass the simulated galaxies have somewhat lower ex-situ mass fractions for all bound stellar mass. Horizon-AGN and Magneticum show much higher ex-situ fraction for low halo masses compared to the other models, while in massive haloes their ex-situ fraction is lower than in \textsc{emerge}, closer to the results of Illustris and TNG. The semi-analytic model agrees well with our predictions, although the semi-analtyic ex-situ fraction is slightly lower than the empirical results.

The ex-situ fraction is very sensitive to the amount of feedback in these models, as shown by \citet{Choi:2015aa} and \citet{Henriques:2019aa}. Because of the shape of the conversion efficiency, feedback in low-mass haloes is stronger in satellites than in centrals. Consequently, if (SN) feedback in low-mass haloes is less efficient, the stellar mass in satellites compared to the stellar mass in centrals is higher, which leads to higher ex-situ fractions. Conversely, in massive haloes feedback is typically stronger in the centrals than in the satellites, such that less efficient (AGN) feedback implies higher stellar mass in the centrals compared to the satellites leading to lower ex-situ fractions. Given this context, it is likely that the SN feedback in Horizon-AGN and Magneticum is comparatively low, such that the stellar mass in the satellites is higher than in the other models. At the massive end, the hydrodynamical simulations have somewhat higher conversion efficiencies than \textsc{emerge}, which is also apparent from their stellar mass functions that are higher for massive galaxies. Consequently, they form more stars in the central galaxies and their ex-situ fractions are slightly lower than our result suggests.

Within the central galaxy only, the ex-situ mass fraction is much lower for each halo mass. Similar as before, below a halo mass of $\log(M_\mathrm{h}/\Msun)=12$, the ex-situ mass fraction is smaller than 1 per cent with very few individual galaxy reaching much higher values. Compared to all bound stars, $f_\mathrm{ex}$ increases less strong with halo mass in the central galaxy (almost linearly with log halo mass), and reaches $\sim30$ per cent at $\log(M_\mathrm{h}/\Msun)=13$ and $\sim60$ per cent at $\log(M_\mathrm{h}/\Msun)=14$. Above $\log(M_\mathrm{h}/\Msun)=15$, the ex-situ fraction in the central galaxy starts to saturate at larger than 80 per cent. We again contrast these results with hydrodynamical simulations including Illustris (magenta line), Magneticum (green line), and TNG (solid blue line: TNG300, dashed blue line: TNG100). However, it is not straight-forward to compare the ex-situ fraction for the central galaxy only. The stellar mass in \textsc{emerge} corresponds to observationally obtained masses (typically using model magnitudes), while in simulations the mass is typically computed within a specific 3D radius. Consequently, a direct comparison is impossible, but we show the results at a typical radius of $30\kpc$. We also show the semi-analytic result for the central galaxy (yellow line), which can be compared to our model directly.

All simulations show significantly higher ex-situ fractions in the central galaxy compared to \textsc{emerge} and the semi-analytic model. The closest agreement is with Illustris, which has only slightly higher ex-situ fractions. For haloes between $\log(M_\mathrm{h}/\Msun)=12$ and $14.5$, the ex-situ mass fractions in TNG are considerably higher than the empirical constraint. Magneticum has an even higher ex-situ fraction, especially for low-mass haloes, where it deviates from all other models. The semi-analytic results again agree quite well with our model, although the ex-situ fraction in the SAM is somewhat lower. At low to intermediate halo masses, the disagreement with the hydrodynamical simulation is again likely caused by more massive accreted satellite galaxies due to less efficient feedback in low-mass galaxies. The second reason why the simulations show higher ex-situ fractions is related to the distribution of the stars in those galaxies, which is relevant for the comparison based on the central stellar mass. Unlike for all bound stellar mass, where it is straight forward to select the star particles, for the central stellar mass it is important up to which radius star particles are selected. In this comparison, the central stellar mass in the simulations has been defined to include all stars within $30\kpc$ of the centre, while for \textsc{emerge} and the semi-analytic model the masses are based on observed model magnitudes. The higher ex-situ fractions in the simulations thus likely imply that the simulation results are based on a larger radius than the empirical constraints.

So far, we have discussed the ex-situ fractions as a function of halo mass, and found that they are close to zero for low-mass haloes with $\log(M_\mathrm{h}/\Msun)<12$, and then rise continuously to over 80 per cent over 3 order of magnitudes in halo mass. Now we investigate more closely, how the ex-situ fractions grow with stellar mass, and determine at which stellar mass accretion becomes the dominant growth channel for galaxies. For this we compute the average ex-situ fraction and the corresponding scatter in bins of stellar mass in the central galaxy at $z=0$, and show the results in the left-hand panel of Figure \ref{fig:macc}. The dots again represent individual systems as computed by \textsc{emerge}, while the black solid line gives the average $f_\mathrm{ex}$ for a fixed central stellar mass at $z=0$. The wide light grey area indicates the $1\sigma$ scatter in this relation, and the narrow dark grey area gives the uncertainty of the average resulting from observational uncertainties as derived the final MCMC walkers. The long-dashed line shows the average ex-situ fraction at fixes central stellar mass at $z=1.0$, and the short-dashed line shows the same quantity at $z=2$. The coloured dotted horizontal lines represent fixed ex-situ fractions of 20, 40, and 60 per cent. In low-mass galaxies, we find that the ex-situ fraction is negligible, while at $\log (m_*/\Msun)=11$ it reaches 20 per cent. This value strongly increases with stellar mass and reaches 40 per cent at $\log (m_*/\Msun)=11.3$, and 60 per cent at $\log (m_*/\Msun)<11.5$. These results are in very good agreement with the conclusions of \citet{Bernardi:2019aa}, who find that mergers start to matter for galaxies with $m_* = 3\times10^{11}\Msun$. At higher redshift, the dependence of the ex-situ fraction on the central stellar mass is very similar, although given values for $f_\mathrm{ex}$ are reached at a slightly higher stellar mass.

To investigate the dependence of the transition mass for which mergers contribute a significant fraction of the stellar mass, we compute the stellar mass at which a given ex-situ fraction is reached as a function of redshift. In the right-hand panel of Figure \ref{fig:macc}, we show the results for six specific ex-situ fractions. The colours of the lines correspond to the same colours in the left-hand panel. We find that for a given ex-situ fraction, the corresponding central stellar mass increases weakly with redshift. A low ex-situ fraction of 10 per cent is reached by galaxies with $\log (m_*/\Msun)=10.9$ at $z=0$, while at $z=1$ galaxies need to have a mass of $\log (m_*/\Msun)=11.1$, and a mass of $\log (m_*/\Msun)=11.2$ at $z=2$. Galaxies for which the stellar mass growth is dominated by accretion, i.e. an ex-situ fraction of more than 50 per cent, need to be more massive than $\log (m_*/\Msun)=11.4$ at $z=0$, and more massive than $\log (m_*/\Msun)=10.6$ at $z=2$. This shows that only the most massive galaxies in the Universe are mainly growing through accretion of satellites, especially at high redshift.


\subsection{Formation of stars in ETGs} \label{sec:sfh}

Having determined through which channels galaxies grow their stellar mass, we now turn to the question of when this mass is formed. In M18, we have shown that massive galaxies have star formation histories that peak at a higher redshift compared to low-mass galaxies. In this section we focus on two aspects: First, when do galaxies in haloes of a fixed mass assemble their stellar mass, i.e. at which redshift does the galaxy reach a certain fraction of its final mass, and second, at what time did the stellar populations in the galaxy form (including the stars that formed ex-situ)?

\begin{figure}
    \includegraphics[width=\halfwidth]{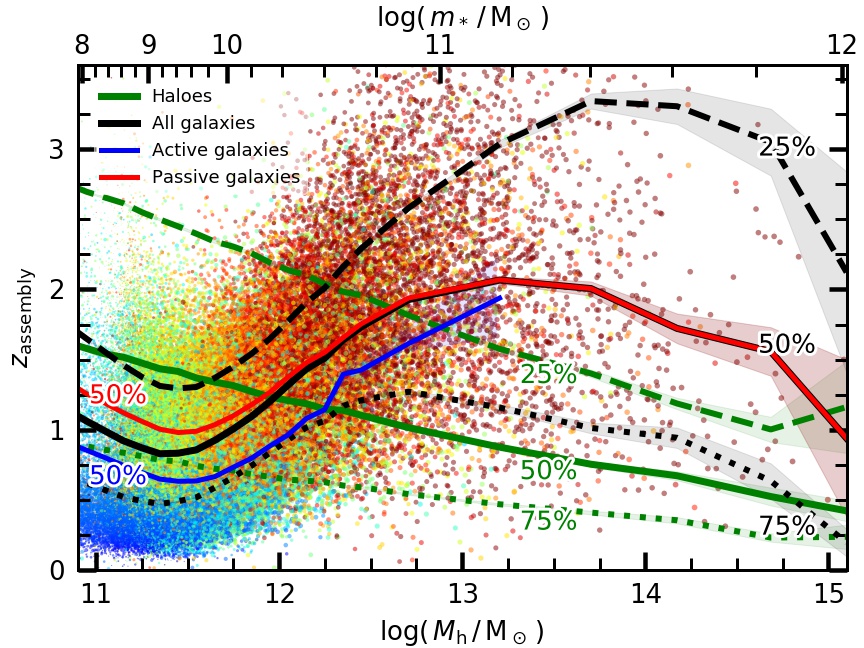}
    \caption{
    Assembly redshift as function of halo mass. Dashed, solid, and dotted lines show the mean redshift at which 25, 50, and 75 per cent of the final mass has assembled, respectively. Green lines show the results for halo mass, while black, blue, and red lines show the results for stellar mass in all, active, and passive galaxies, respectively. The dots in the background show the 50 per cent assembly redshift of individual galaxies, using the specific SFR colour-coding of Figure \ref{fig:shm}. The top axis indicates the stellar mass with an average halo mass that corresponds to the bottom axis.
    }
    \label{fig:zassembly}
\end{figure}

\subsubsection{Stellar mass assembly} \label{sec:zassembly}

To address the first issue, we compute the redshift at which the central galaxy in haloes of a fixed mass has assembled a given fraction of its final stellar mass for the first time. In this context, assembly means that the stars are part of the central galaxy, but can have formed either within the central galaxy or in an accreted satellite galaxy. We show the results in Figure \ref{fig:zassembly} for 25 (dashed lines), 50 (solid lines), and 75 per cent (dotted lines). The mean assembly redshift for all galaxies are given by the black lines, while the mean 50 per cent assembly redshift for active and passive galaxies are given by the blue and red line, respectively. For comparison, we also show the mean redshift at which the dark matter haloes have assembled 25, 50, and 75 per cent of its final mass (green lines). The points in the background show the 50 per cent assembly redshift for individual galaxies, colour-coded by specific SFR (c.f. Figure \ref{fig:shm}).

While low-mass haloes tend to assemble their mass relatively early -- 50 per cent of the final halo mass of $10^{11}\Msun$ has assembled by $z=1.5$ -- the stellar mass in the central galaxy is in place only relatively late -- 50 per cent of the final stellar mass has assembled by $z=1$. Conversely, in more massive haloes the final halo mass assembles late ($z=1$ for 50 per cent in haloes of $10^{13}\Msun$), while the final stellar mass is in place much earlier ($z=2$ for 50 per cent). This effect is sometimes called `anti-hierarchical' galaxy formation, and results from the shape of the instantaneous conversion efficiency. While low-mass haloes are already in place at high redshift, the conversion of gas into stars is inefficient until the halo has gained enough mass. Only then will the galaxy be able to form stars efficiently, resulting in a late assembly redshift. In more massive haloes, this situation is reversed, i.e. while the haloes get most of their mass late, they have passed through the peak of the conversion efficiency at higher redshift, such that the central galaxy was able to form stars efficiently. Consequently, most of the stellar mass is forming early, while at late times, the haloes become massive enough to have a low conversion efficiency, such that the galaxies are quenched and only very little stellar mass assembles late, practically only through merging satellites.

\begin{figure*}
    \includegraphics[width=\halfwidth]{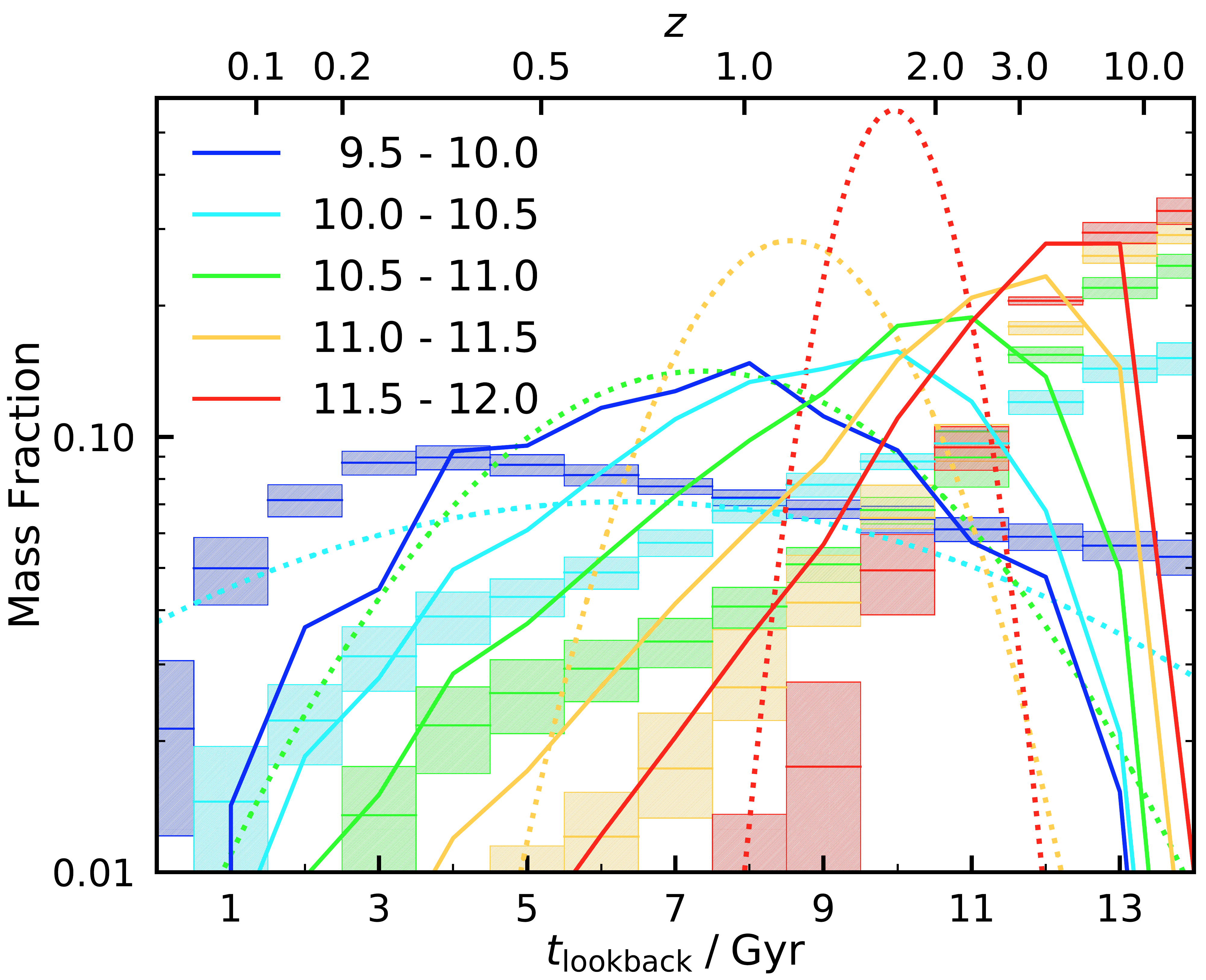}
    \hspace{5mm}
    \includegraphics[width=\halfwidth]{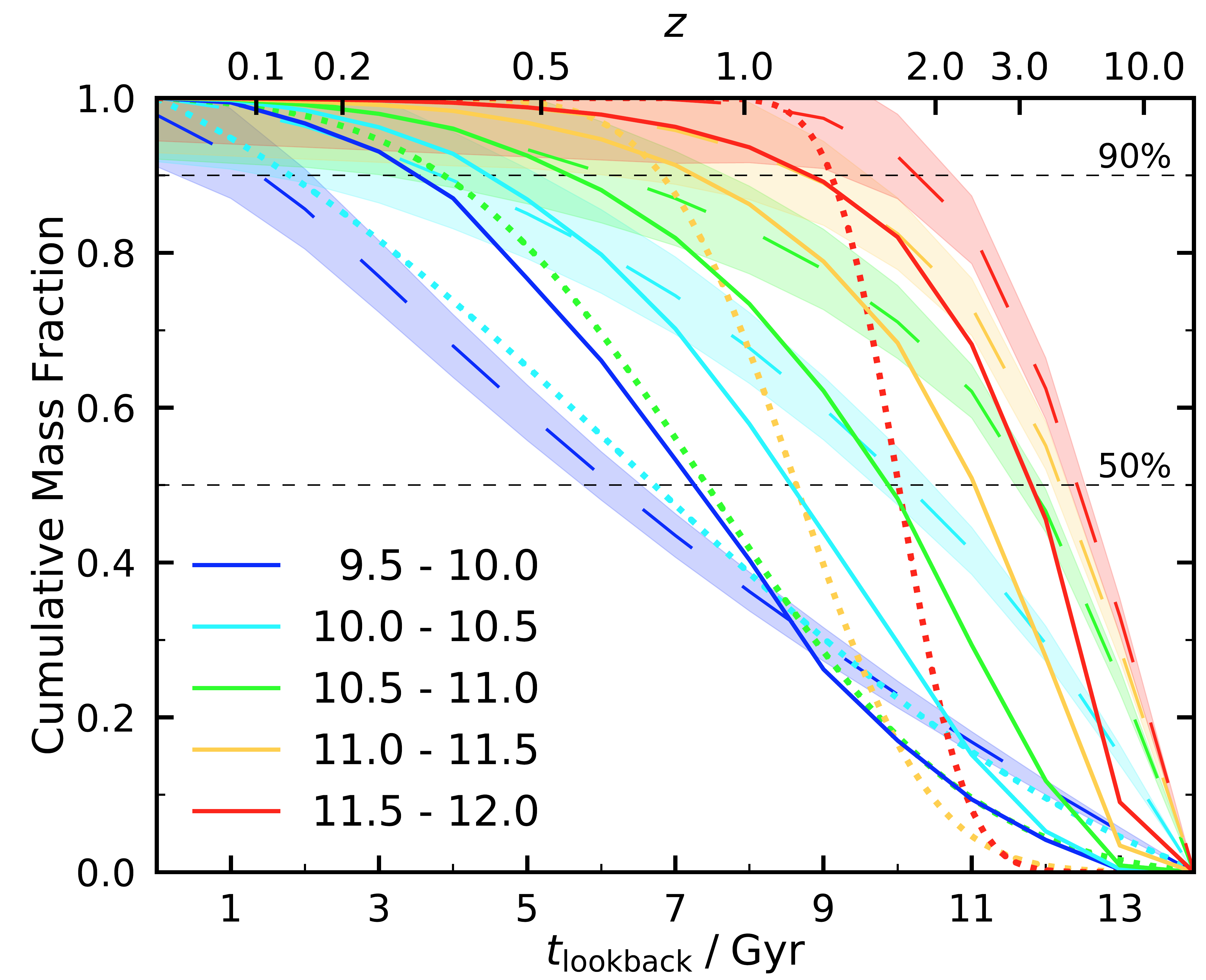}
    \caption{
    \textit{Left-hand panel:} The fraction of the stellar mass in passive galaxies at $z=0$ which has formed at a given lookback time in bins of 1 \Gyr. The solid lines with different colours show the results of \textsc{emerge} for 5 stellar mass bins, the coloured dotted lines show the results by \citet{Thomas:2010aa} based on SDSS, and the coloured rectangles indicate the results of the ATLAS$^{\rm 3D}$ survey \citep{McDermid:2015aa}.
    \textit{Right-hand panel:} Cumulative mass fraction as a function of lookback time, indicating which fraction of the present-day stellar mass has formed before a specific lookback time. The solid lines again show the results of \textsc{emerge}, the dotted lines show the SDSS results, and the dashed lines with the shaded area show the ATLAS$^{\rm 3D}$ observations.
    The stars in massive galaxies form at very early cosmic times, while the stellar mass in low-mass galaxies forms much later. As all galaxies in these samples are passive, the fraction of the stellar mass formed at late times ($t_\mathrm{lb} < 5\Gyr$) is very low.
    }
    \label{fig:massfractions}
\end{figure*}

In very massive haloes though, the situation changes again, as the assembly is increasingly dominated by merging satellites (see section \ref{sec:inexsitu}). While star formation still mainly happens at high redshift, satellites are predominantly accreted late, so that the resulting assembly redshifts are shifted to lower values. This also explains why the mean 25 per cent assembly redshift is still very high ($z>3$ in haloes of $\sim10^{14}\Msun$), but the mean 75 per cent assembly redshift is considerably lower (at $z < 0.5$). Independent of halo halo mass, active galaxies assemble later than passive galaxies at fixed halo mass, which simply follows from the enhanced SFRs of active galaxies at late times compared to passive galaxies.

\subsubsection{Star formation histories of ETGs} \label{sec:massfractions}

Instead of looking at the time when stellar mass finally assembles in ETGs, we now turn towards the formation of the stars in the first place. For this we select all passive galaxies at $z=0$ and divide this sample into five stellar mass bins. We then take all stellar mass that is still in place at $z=0$ (and has not been lost by dying stars), and compute the fraction that has formed in a certain lookback time bin of width $\Delta t = 1\Gyr$. The results are shown in the left-hand panel of Figure \ref{fig:massfractions} for all five mass bins (solid lines). The right-hand panel of Figure \ref{fig:massfractions} shows the cumulative mass fractions as a function of lookback time for the same mass bins, i.e. which fraction of the stars present at $z=0$ has formed before a specific lookback time. This shows that in massive galaxies, most of the stars that are still present today have formed at early cosmic times. For the most massive galaxies with $\log (m_*/\Msun)$ between 11.5 and 12.0 about 30 per cent of the remaining stars form between a lookback time of 12.5 and $13.5 \Gyr$, and another 30 per cent between 11.5 and $12.5 \Gyr$. In the right-hand panel, we can identify that 50 per cent of the stars in this mass bin have formed before a lookback time of $11.8 \Gyr$, i.e. within the first $2 \Gyr$ of the Universe.

The lower the stellar mass of the galaxies, however, the later its stars formed. Intermediate-mass galaxies with $\log (m_*/\Msun)$ between 10.5 and 11.0 form most of their stars at a lookback time of around $10 \Gyr$, which corresponds to the time when half of their present stellar mass had been formed. For low-mass galaxies with $\log (m_*/\Msun)$ between 9.5 and 10.0 the bulk of the star formation happens at a lookback of $8 \Gyr$ and half of their $z=0$ stellar mass has formed before a lookback time of $7.3 \Gyr$. As all galaxies in the samples are passive, the fraction of the stellar mass formed at late times ($t_\mathrm{lb} < 5\Gyr$) is generally very low.

\begin{figure*}
    \includegraphics[width=\fullwidth]{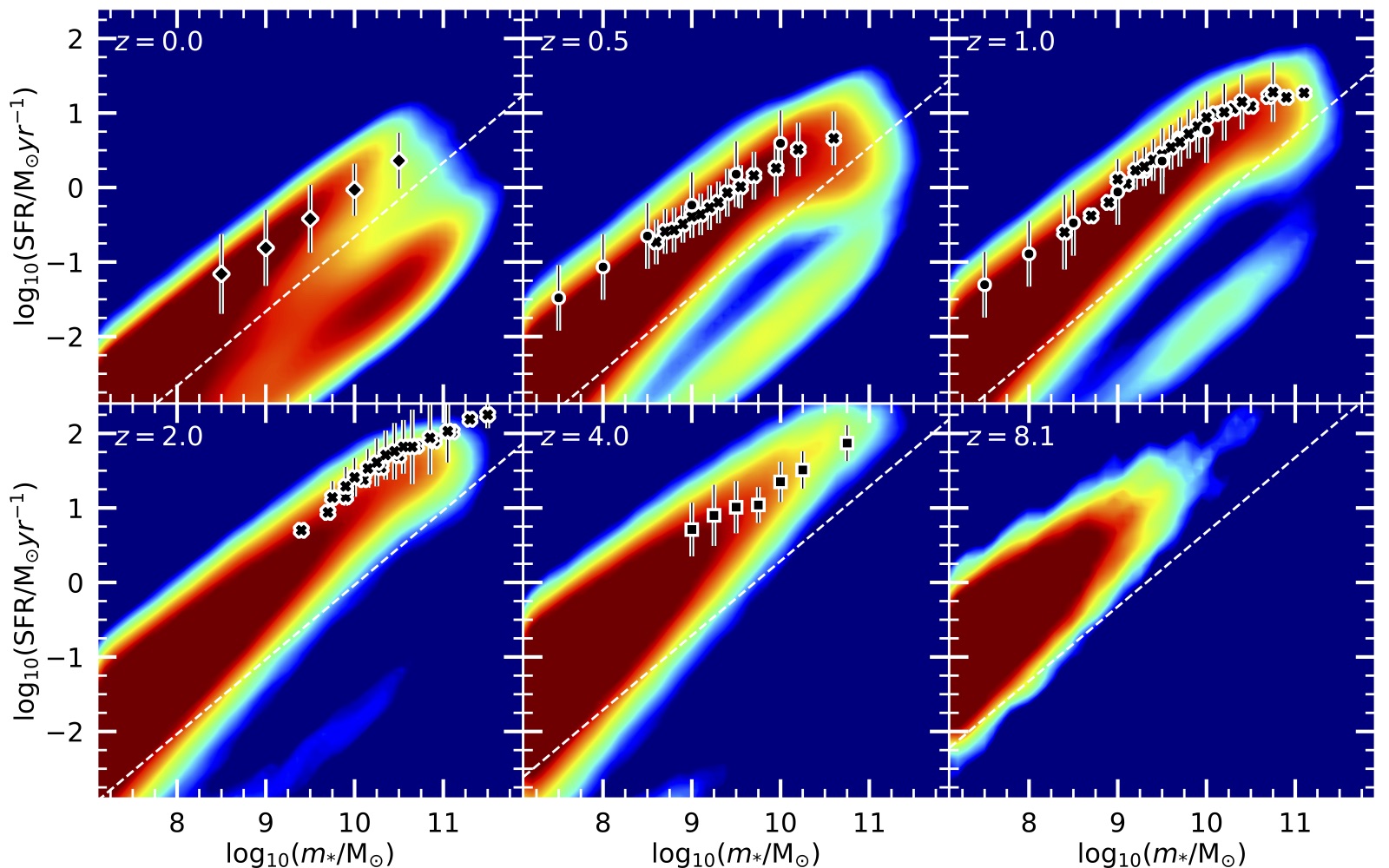}
    \caption{
    Model prediction for the dependence of the star formation rate on the stellar mass at six redshifts from $z=0$ to $z=8$ (from top left to bottom right). The colour of the $\mathrm{SFR}-m_*$ map indicates how many galaxies can be found in any location, i.e. there are many galaxies in the red areas and no galaxies in the blue areas. The white dashed lines corresponds to the specific SFR threshold that separates active and passive galaxies at each redshift. The symbols with error bars show the results found by surveys: diamonds for SDSS \citep{Renzini:2015aa}, crosses for 3D-HST \citep{Whitaker:2012aa, Whitaker:2014aa}, circles for MUSE-HUDF \citep{Boogaard:2018aa}, and squares for CANDELS \citep{Salmon:2015aa}.
    }
    \label{fig:sfrz}
\end{figure*}

Galactic archeology allows to derive the formation times of stellar populations by observational means directly. In Figure \ref{fig:massfractions}, we compare our results to formation times obtained observationally by \citet{Thomas:2010aa} based on SDSS (dotted lines), and by \citet{McDermid:2015aa} based on the ATLAS$^{\rm 3D}$ survey (rectangles and dashed lines). The mean stellar ages in the SDSS data have been obtained from Lick absorption line indices, while the formation time scales are based on the measured $\alpha$/Fe element ratios. These parametrised results show the same trend for higher mass galaxies forming their stars at higher redshift but miss the extended histories for towards lower redshift for low mass galaxies. The star formation histories in the ATLAS$^{\rm 3D}$ data have been derived by fitting a linear combination of single stellar populations (SSPs) model spectra to the observed galaxy spectra integrated within apertures covering up to one effective radius. To deal with the inherent degeneracy of inferring the star formation history from integrated light, linear regularisation is used such that the resulting star formation history is as smooth as possible while still reproducing the observed spectrum. The observationally derived star formation histories agree fairly well with our model predictions. However, we find that the observed data is somewhat more extreme than our model. Massive galaxies form even earlier in the observations, while low-mass galaxies tend to have even later star formation compared to our model. One difficulty in this comparison is obviously the sample selection. While the ATLAS$^{\rm 3D}$ sample has been selected morphologically as ETGs, we have selected our sample as passive galaxies with a cut in the specific SFR. This means that wile the observed sample still mainly consists of passive galaxies, it will contain a fraction of active galaxies, especially at the low-mass end. As active galaxies have much younger stellar populations than passive galaxies, these few active galaxies in the sample cause the fraction of stars formed at late cosmic times to increase, while at the same time the relative fraction of stars formed at early times decreases. This effect accounts for most of the difference between modelled and observationally inferred star formation histories for low and intermediate-mass galaxies.

In summary, our empirical model \textsc{emerge} implies that the most massive galaxies have the oldest stellar populations, i.e. they have star formation histories that  peak at very early cosmic times. In contrast, galaxies of lower mass contain younger stellar populations as their star formation histories peak later. This `downsizing' effect is a result of the shape of the baryon conversion efficiency, as massive haloes and galaxies are most efficient in converting baryons to stars at early cosmic times when their halo reaches a mass of around $10^{11.5}\Msun$, while low-mass haloes reach this most efficient mass only at late cosmic times. However, if we do not look at when the stars were formed in the first place, but when they assemble in the final galaxy, we find that the most massive galaxies have rather late assembly times. This can be explained by stellar growth in massive galaxies being dominated by ex-situ star formation. While the vast majority of the stars in massive galaxies have formed very early, a large fraction of those stars formed in other galaxies and has only been accreted onto the final galaxy at a later time. Therefore, the galaxies that have assembled most of their stars the earliest are not the most massive galaxies, but those that live in haloes of $\log (M_\mathrm{h}/\Msun) \approx 13$, and have typical stellar masses of $\log (m_*/\Msun) \approx 11$. Independent of this, we find that at a fixed halo mass, passive galaxies form and assemble earlier than active galaxies.



\subsection{The build-up of the star-formation bimodality} \label{sec:starformation}

So far we have investigated how galaxies build up their mass and found that massive galaxies typically form their stars at relatively early cosmic times. Comparing our results to observations that have not been used in constructing the empirical model, such as the SHM relation for active and passive galaxies from lensing and the star formation histories in ATLAS$^{\rm 3D}$, we obtained a very good agreement. Now we check if the predictions by \textsc{emerge} are able to meet other independent observational constraints on the SFR of galaxies. For this we determine how the SFR depends on stellar mass at different redshifts and confront this with observations from major surveys. Once these constraints have been met, we investigate how the progenitors of passive galaxies follow these scaling relations and when they start to depart from it.

\subsubsection{The relation between stellar mass and star formation rate} \label{sec:sfrz}

Before we can compare the relation between stellar mass and SFR to observations, we need to define an `observational SFR' for model galaxies that have extremely low or zero SFRs. In practice, these galaxies would not have any signal in the bands that are typically used to derive observed SFRs. To still be able to show these galaxies in observational data, it is common to set their SFRs to the maximum possible value, i.e. the detection limit of the survey. We simulate this effect by adopting a minimal specific SFR before we apply any observational uncertainty. We apply a minimal specific SFR of $10^{-12}\mathrm{yr}^{-1}$ and a scatter of $0.3\mathrm{dex}$ to derive our mock SFR from the model SFR. In Figure \ref{fig:sfrz}, we show the relation between SFR and stellar mass derived with \textsc{emerge}. The colour scale indicates how many galaxies with a given stellar mass-to-SFR combination can be found, i.e. the red areas are populated by many galaxies, while there are no galaxies in the blue regions of the plot. The white dashed lines represents the specific SFR threshold used to divide active and passive galaxies at each redshift.

At high redshift of $z\sim8$, we observe that the star formation main sequence is building up. At this cosmic time, the relation is still very wide with a large scatter, resulting from highly fluctuating SFRs of individual galaxies. At later times, these fluctuations become smaller and the main sequence becomes tighter. At $z\sim2$, the main sequence is fully in place and comprises the vast majority of all galaxies. At this redshift, we observe two effects. The first effect is the bending of the main sequence at high stellar masses towards lower SFRs. This downturn can be explained by the onset of feedback in massive central galaxies. The main sequence follows a power-law up to the stellar mass where the conversion efficiency starts to decrease again, which corresponds to the characteristic mass (`knee') of the SMF. Here, the conversion of baryons into stars becomes less efficient, resulting in lower SFRs. At $z\sim2$, many galaxies cross this stellar mass, such that the bend in the main sequence becomes apparent. We tested an alternative implementation of the conversion efficiency in \textsc{emerge}, where instead of gradually decreasing with halo mass, the conversion efficiency instantaneously drops to zero once a threshold halo mass has been crossed. In this model we do not find any bending in the main sequence, indicating that this effect is caused by a gradual decline of the conversion efficiency, i.e. feedback becomes stronger with halo mass instead of acting above a certain halo mass. The second effect we notice, is the emergence of a cloud of passive galaxies after $z\sim2$. In \textsc{emerge}, these are galaxies living in haloes that have stopped growing, such that the galaxies have eventually quenched. Most of these galaxies are satellites.

Towards lower redshift, we observe that the bend at the massive end of the main sequence becomes more apparent. While the haloes of those galaxies still grow, their conversion efficiency becomes much lower. Consequently, their SFRs drop and their stellar mass does no longer grow, which results in lower SFRs at fixed stellar mass, and an even stronger bend in the main sequence. At the same time, more galaxies get instantly quenched in haloes that have stopped growing (or started to lose mass). As a result, the passive cloud becomes more prominent towards low redshift. Finally, at $z=0$, the conversion of gas into stars in massive galaxies is so inefficient that the SFRs in those galaxies are extremely low, so that they are also located in the passive cloud. This area has become much more prominent at $z=0$, and includes centrals as well as satellites. The bend in the main sequence has moved to lower stellar masses, while at the same time there remain many more massive galaxies on the main sequence that no longer accrete new baryonic material, but convert the baryons that have already accreted.

We compare these results to the observational findings of SDSS \citep[diamonds]{Renzini:2015aa}, 3D-HST \citep[crosses]{Whitaker:2012aa, Whitaker:2014aa}, MUSE-HUDF \citep[circles]{Boogaard:2018aa}, and CANDELS \citep[squares]{Salmon:2015aa} in Figure \ref{fig:sfrz}. At high redshift the model is in excellent agreement with the CANDELS observations. At intermediate redshift ($z=1-2$), the model agrees well with the observational constraints, although the observations tend to find slightly higher values for the SFR on the main sequence. The model prediction for the bending of the main sequence around $\log(m_*/\Msun) \ge 10$ is in very good agreement with the observations. This indicates that the conversion efficiency at the massive end has to decline gradually with halo mass instead of instantaneously dropping above a certain halo mass, and that feedback at the massive end acts more strongly with increasing halo mass. At low redshift, we find very good agreement with SDSS and 3D-HST. As \textsc{emerge} is able to reproduce these individual constraints up to high redshifts, we will use the predictions for the SFRs in combination with the information how different galaxy populations are linked through cosmic time to further investigate the progenitors of ETGs.

\begin{figure}
    \includegraphics[width=\halfwidth]{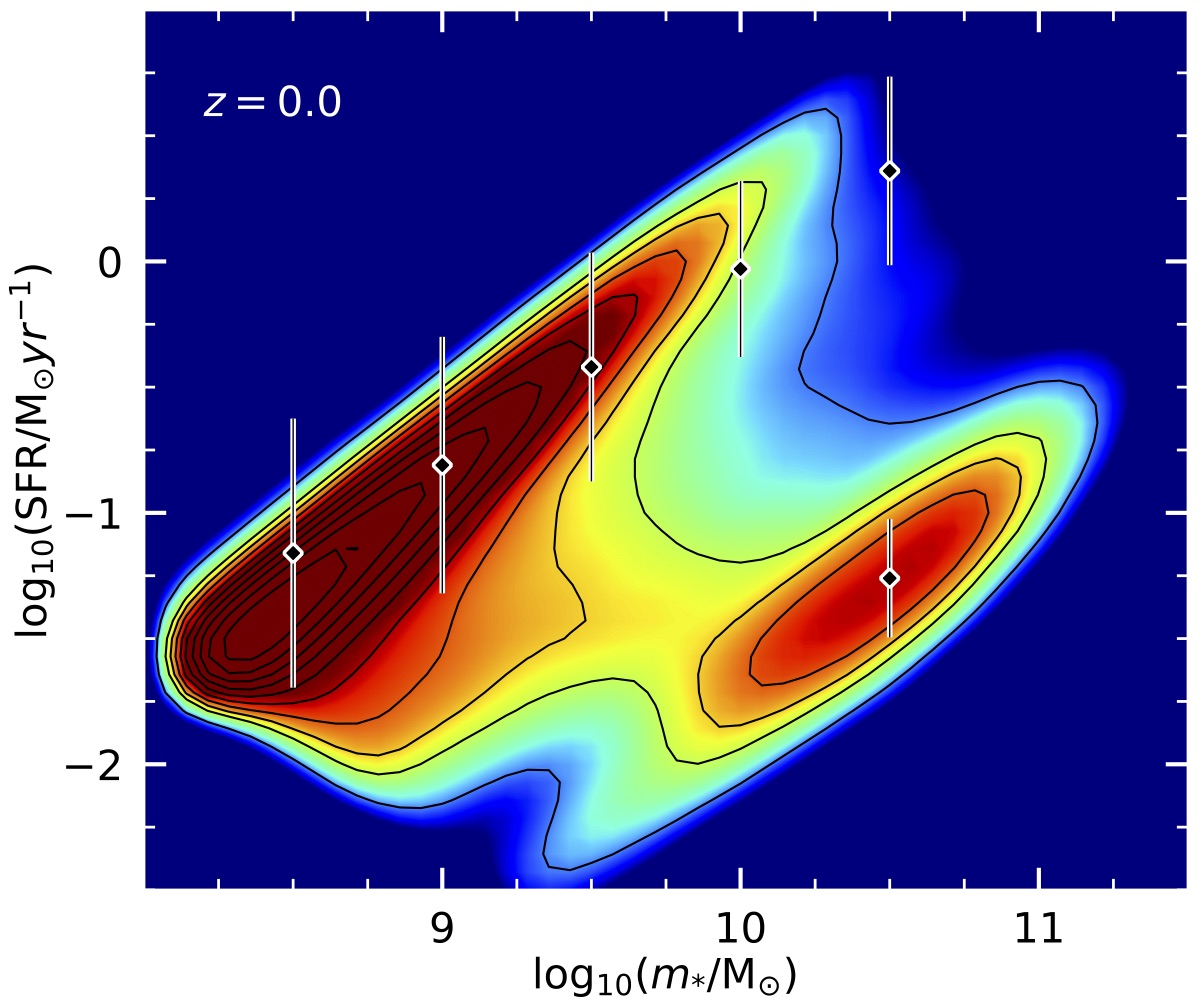}
    \caption{
    Mock observation for relation between stellar mass and star formation rate at $z=0$. The colour indicates how many galaxies can be found in a given region. The map has been derived by applying a minimum specific SFR of $10^{-12}\mathrm{yr}^{-1}$, observational uncertainty of $0.3\mathrm{dex}$, and a combined completeness limit for stellar mass and star formation rate. The diamonds correspond to the results from SDSS \citep{Renzini:2015aa}, and the error bars indicate the $1\sigma$ scatter in the observed relations.
    }
    \label{fig:sfr0}
\end{figure}

Before we study when and how passive galaxies were quenched and moved onto the red cloud, we focus on the relation between stellar mass and SFR for the present galaxy population, and compare our results to observations. Figure \ref{fig:sfr0} presents a mock observation for local galaxies obtained with \textsc{emerge} by applying the previously stated minimum specific SFR and observational uncertainty. Further, to take into account the completeness limits of SDSS, we remove all galaxies for which the product of the stellar mass and SFR is lower than $10^{6.5}\Msun^2\mathrm{yr}^{-1}$, such that galaxies with a low stellar mass are only included in the plot if they have a relatively high SFR. We further show the SDSS results by \citet{Renzini:2015aa} as diamonds (c.f. their Figure 4), both for the star formation main sequence and the passive cloud. The error bars correspond to the observed $1\sigma$ scatter in those relations. We find a very good agreement between data and model predictions, although we note that the slope in the observed main sequence is slightly flatter. At this stage it remains unclear, whether this discrepancy results from a bias in the data, or from a deficiency in the model. The passive cloud is centred around $\log(m_*/\Msun)=10.5$ and  $\log(\Psi/\Msun\mathrm{yr}^{-1})=-1.3$, both in the data and in the model.

\subsubsection{Star formation in progenitors of passive galaxies} \label{sec:zquenching}

Having verified that \textsc{emerge} predicts SFRs that are in agreement with observational constrains up to high redshift, we now focus on the build-up of the star-formation bimodality and investigate how massive galaxies end up on the passive cloud at $z=0$. For this we divide all model galaxies including satellites in the present-day Universe into an active and a passive sample, and mainly concentrate on the latter. We then trace them through cosmic time and study the properties of their progenitors, and how they evolve with redshift.

\begin{figure}
    \includegraphics[width=\halfwidth]{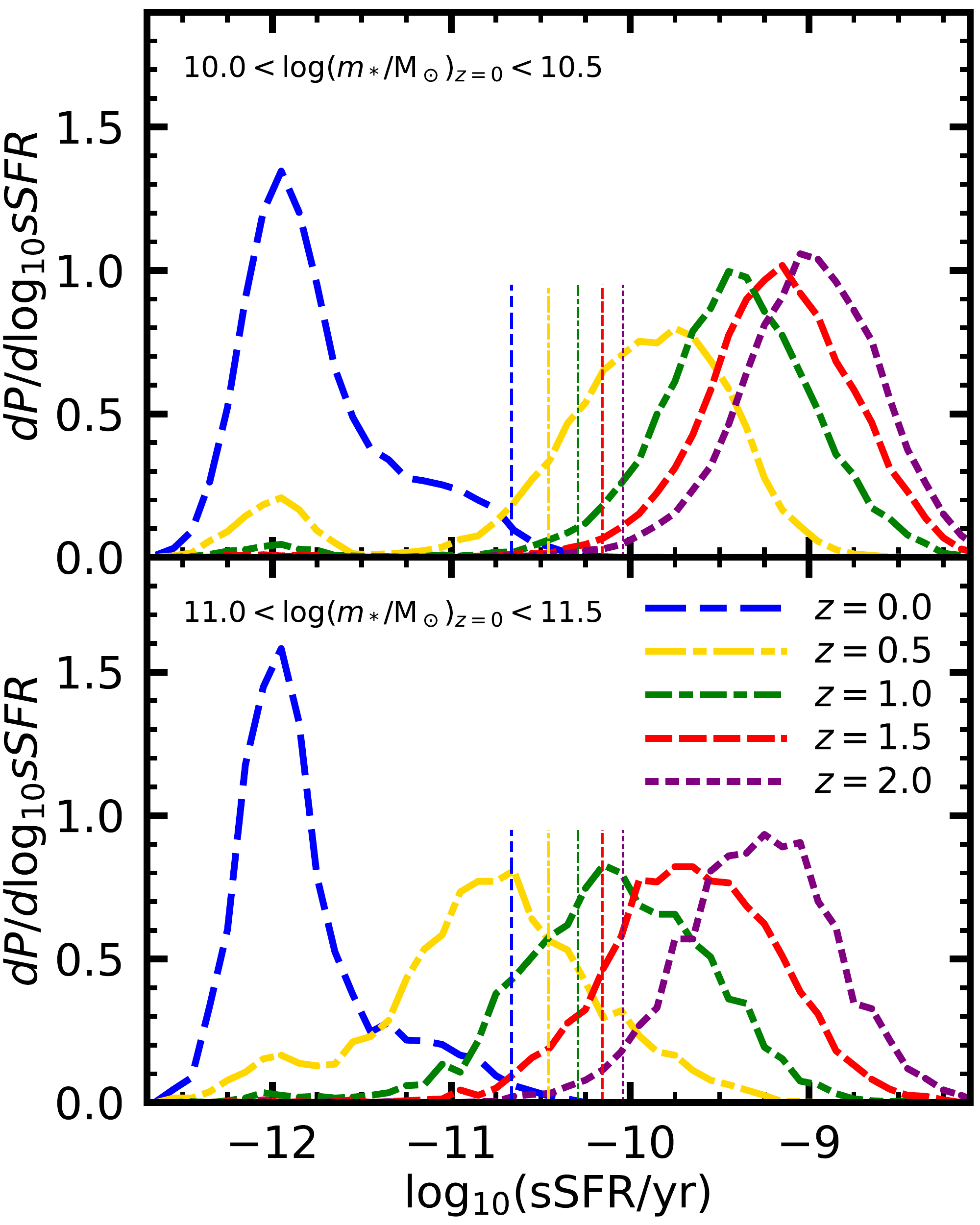}
    \caption{
    Distribution of specific SFRs of the main progenitors of presently passive galaxies. The dashed lines with different colours show the specific SFR distributions for fixed progenitor redshifts. The thin vertical lines indicate the threshold that distinguishes active from passive galaxies for each redshift. The top panel gives the results for intermediate mass galaxies, and shows that almost all of their progenitors have high SFRs before $z=1$. The bottom panel gives the results for massive galaxies, and shows that their progenitors have almost exclusively high SFRs only before $z=2$.
    }
    \label{fig:progenitorSSFR}
\end{figure}

\begin{figure}
    \includegraphics[width=\halfwidth]{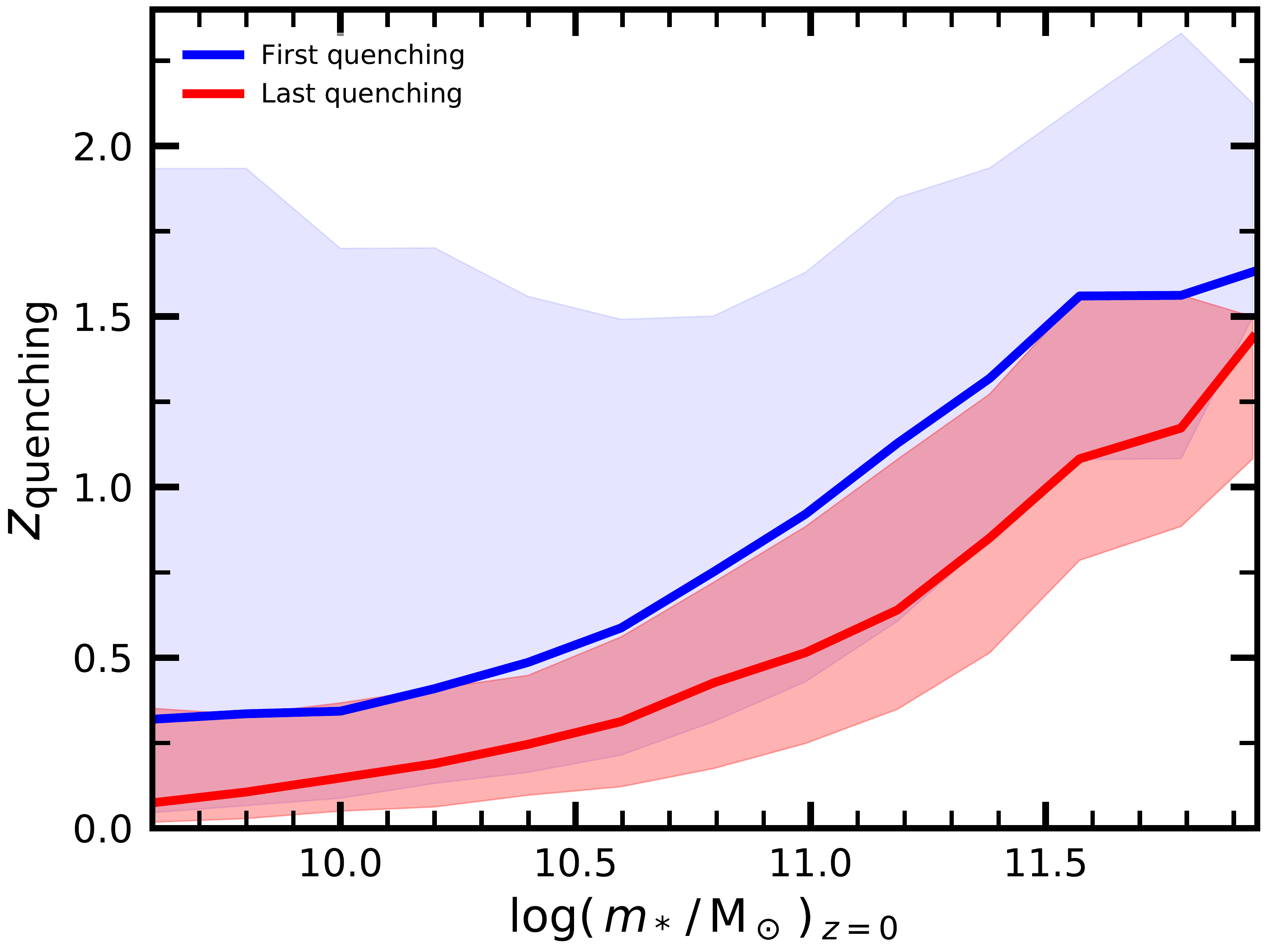}
    \caption{
    Quenching redshift of presently passive galaxies as a function of their $z=0$ stellar mass. The blue and red lines show the average redshift, when galaxies are quenched for the first and last time in their history, respectively. The shaded areas correspond to the $1\sigma$ scatter in this relation. The majority of massive galaxies become finally quenched before $z=1$, while low-mass galaxies are only quenched at very low redshift.
    }
    \label{fig:zquenching}
\end{figure}

In Figure \ref{fig:progenitorSSFR}, we show the distribution of specific SFRs of the main progenitors of presently passive galaxies. To derive these results, we select all passive galaxies in a specified stellar mass bin at $z=0$ and follow their main branches (tracing the most massive progenitor) up to the indicated redshifts. We then compute the specific SFR for each main progenitor and derive the probability density function at any redshift. These specific SFR distributions are plotted for each redshift (coloured lines) for intermediate $z=0$ stellar masses of $10.0\le\log (m_*/\Msun)<10.5$ in the upper panel and massive galaxies with $11.0\le\log (m_*/\Msun)<11.5$ in the lower panel. As above, we have applied a minimum specific SFR of $10^{-12}\mathrm{yr}^{-1}$ and an observational uncertainty of $0.3\mathrm{dex}$. The dotted vertical lines indicate the threshold that divides active and passive at each redshift.

\begin{figure}
    \includegraphics[width=\halfwidth]{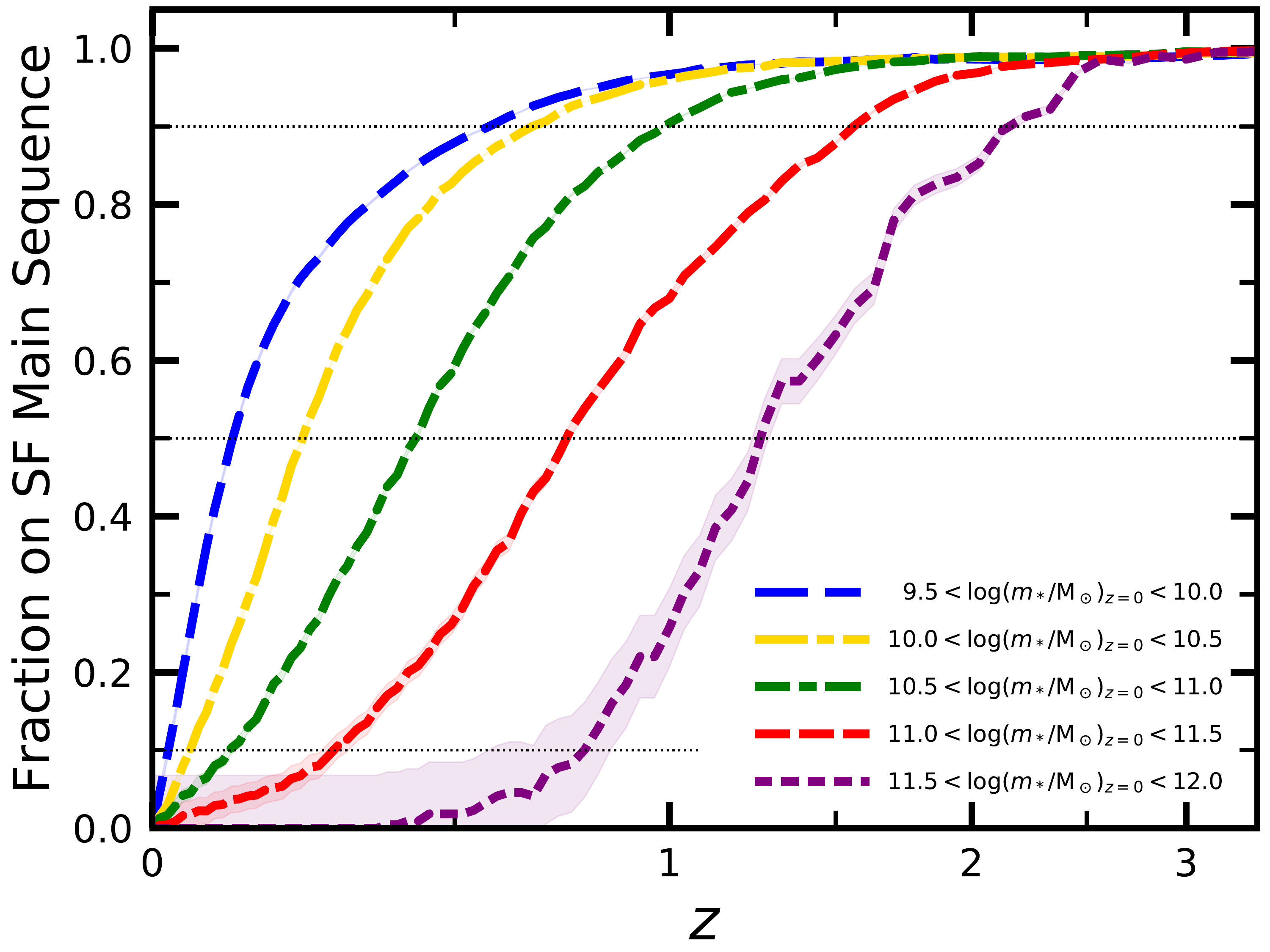}
    \caption{
    Fraction of main progenitors of presently passive galaxy that are on the star formation main sequence at a given redshift. The dashed lines with different colours show the results for different $z=0$ stellar mass bins. The vast majority (90 per cent) of the main progenitors of low-mass presently passive galaxies lie on the main sequence before $z=0.5$, while for the most massive presently passive galaxies the fraction of main progenitors on the main sequence is only 90 per cent before $z=2$.
    }
    \label{fig:progenitorFQ}
\end{figure}

The specific SFRs of the selected galaxies at $z=0$ are low and centred around the adopted minimum value, as we have only included passive galaxies. Consequently, all galaxies have specific SFRs below the division threshold. We note that intermediate mass sample shows a slightly wider distribution for the passive mode compared to the massive sample. The reason for this are galaxies that have recently departed the main sequence and have not yet fully settled in the passive cloud, i.e. their SFRs are still higher than the minimum value. In contrast, the massive sample includes much fewer of these recently quenched galaxies and therefore shows a narrower distribution. The specific SFR distributions of the main progenitors move towards higher values with higher redshift. The progenitors of the intermediate mass sample still have a small mode of quenched galaxies around the minimum specific SFR value at $z=0.5$. The vast majority of the progenitors are actively forming stars with the mode  centred around a specific SFR of $\log (\mathrm{sSFR}/\mathrm{yr}^{-1})=-9.8$. At higher redshift this mode further moves to higher specific SFR values and reaches $\log (\mathrm{sSFR}/\mathrm{yr}^{-1})=-9.0$ at $z=2$. At the same time, there are no passive progenitors, so that we can conclude that all progenitors of intermediate mass passive galaxies lie on the star formation main sequence at $z\sim1.5$.

In contrast, massive galaxies are passive and have low specific SFRs much earlier. The progenitors of the massive sample are mostly already passive at $z=0.5$ with the mode of the distribution at $\log (\mathrm{sSFR}/\mathrm{yr}^{-1})=-10.8$. Only at $z=1$, more than half of the main progenitors are active and have specific SFRs centred around $\log (\mathrm{sSFR}/\mathrm{yr}^{-1})=-10.1$. The mode of the specific SFR moves to higher values with increasing redshift, although the peak is at lower values for the progenitors of the massive sample compared to the progenitors of the intermediate mass sample, which is again a manifestation of the bending in the star formation main sequence. At $z=2$ almost all progenitors of the massive sample are active, and at $z\sim2.5$, all they all lie on the star formation main sequence.

Using these results, we can compute at which redshift the galaxies in the passive sample become quenched. As galaxies usually do not have monotonic star formation histories, they can be quenched multiple times with periods of active star formation in between. Therefore, we compute the the redshift at which passive galaxies at $z=0$ were quenched for the first time, and the redshift when they were quenched for the last time in their history. We show the resulting quenching redshifts in Figure \ref{fig:zquenching} as a function of the stellar mass at $z=0$. The blue and red lines correspond to the peak in the distribution of the first and last quenching redshifts at fixed $z=0$ stellar mass, respectively. The blue and red shaded areas show the upper and lower $1\sigma$ values for both distributions, i.e. 68 per cent of all passive galaxies are quenched within this interval.

\begin{figure}
    \includegraphics[width=\halfwidth]{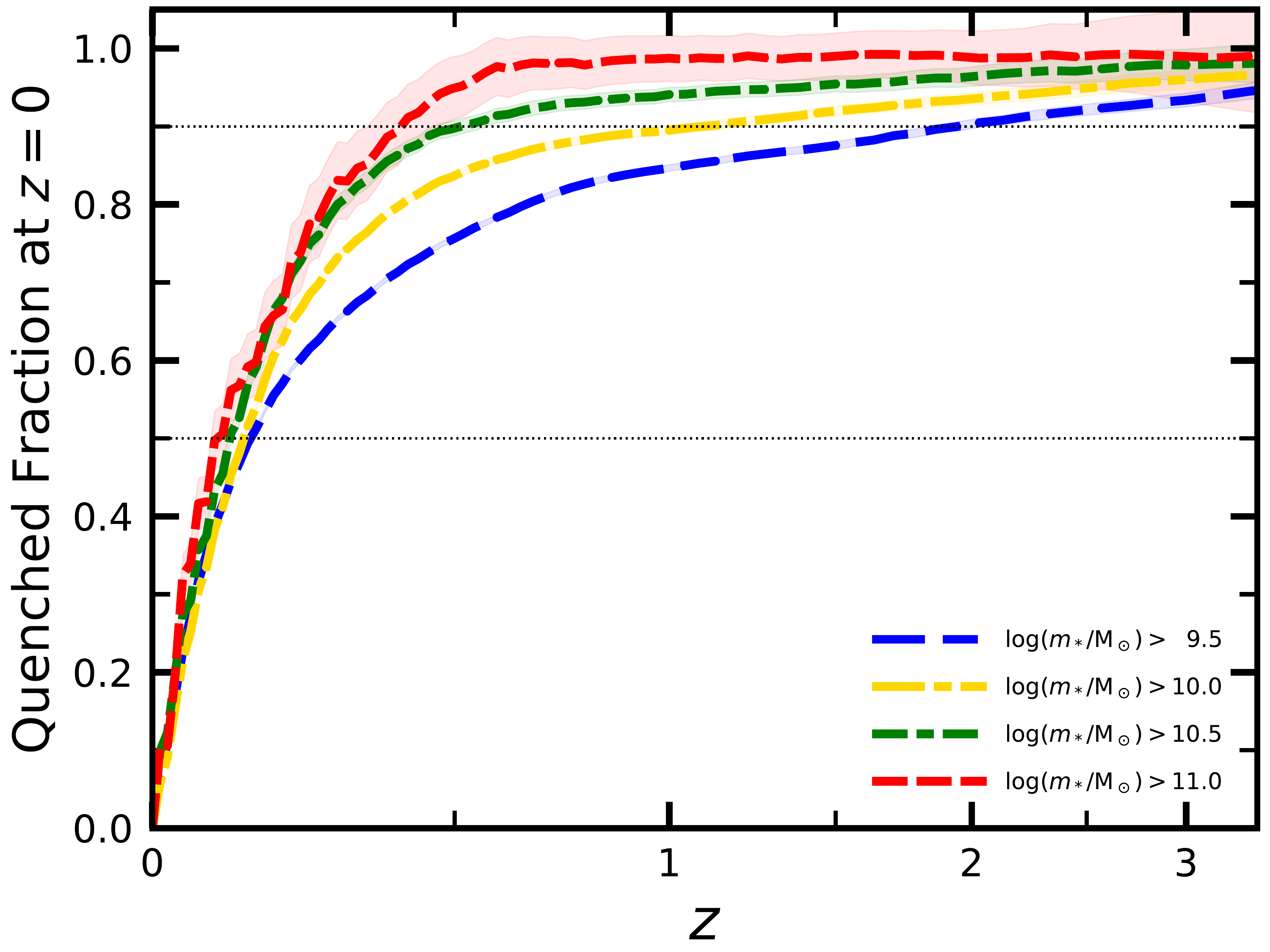}
    \caption{
    Fraction of main sequence galaxies at redshift $z$ that become quenched by $z=0$. The dashed lines with different colours show the results for different stellar mass thresholds at the indicated redshift. Over 90 per cent of the main sequence galaxies with $\log (m_*/\Msun)>9.5$ at $z=2$ become quenched by $z=0$, while only 75 per cent of the main sequence galaxies with $\log (m_*/\Msun)>9.5$ at $z=0.5$ are passive today. Virtually all main sequence galaxies with $\log (m_*/\Msun)>11$ at $z=2$ are quenched today, and still 95 per cent of main sequence galaxies with $\log (m_*/\Msun)>11$ at $z=0.5$ have left the main sequence by now.
    }
    \label{fig:descendantFQ}
\end{figure}

Low-mass galaxies are typically quenched very late. While galaxies with a stellar mass of $\log (m_*/\Msun)=9.5$ can be quenched for the first time as early as $z=2$, most of them are quenched for the first time only by $z=0.3$, and very few of them after $z=0.1$. In contrast, the redshift when those galaxies are quenched for the final time is much lower. Some galaxies were finally quenched by $z=0.6$, while the majority became passive by $z=0.1$, and a some were only quenched very recently. Intermediate mass galaxies with $\log (m_*/\Msun)\sim10.75$ tend to have higher quenching redshifts than low-mass galaxies. While the first of them are quenched a little later around $z=1.5$, the majority are quenched for the first time before $z=0.75$, and almost all of them went through a passive episode by $z=0.3$. The final quenching redshift for this sample is typically around $z=0.4$, while most of them became fully passive by $z=0.2$. The most massive galaxies with stellar masses of $\log (m_*/\Msun)=12$ were quenched very early. Quenching can start as early as $z=2.5$ and most galaxies have undergone a passive period by $z=1.7$. Most progenitors have been completely passive since $z=1.5$, and only very few of them have formed stars after $z=1$.

This analysis shows that before $z\sim2$ almost all progenitors of presently passive galaxies have continuously formed stars and were located on the main sequence. The most massive galaxies have departed from the main sequence a long time ago (at $z\sim1.5$) and were passive ever since, while galaxies with a lower stellar mass at $z=0$ stayed on the main sequence much longer, and only left it rather recently (after $z\sim0.5$). To quantify this behaviour in more detail, we compute the fraction of all main progenitors of presently passive galaxies that are still on the main sequence. For this we select all passive galaxies within a $z=0$ stellar mass bin and record how many of their main progenitors are still actively forming stars at any given redshift, having specific SFRs that are higher than the threshold. We show the resulting fractions as a function of redshift in Figure \ref{fig:progenitorFQ} for five $z=0$ stellar mass bins (dashed coloured lines).


The fraction of main progenitors of low-mass passive galaxies with a $z=0$ stellar mass of $9.5\le\log (m_*/\Msun)<10.0$ that are on the main sequence rises quickly with redshift. At $z=0.1$, already 50 per cent of the progenitors are located on the main sequence. At $z=0.5$, 90 per cent of the progenitors of low-mass passive galaxies are on the main sequence, and at $z=1.3$ virtually all (99 per cent) of the progenitors are forming stars. Intermediate-mass galaxies with a $z=0$ stellar mass of $10.5\le\log (m_*/\Msun)<11.0$ leave the main sequence earlier. Half of their progenitors have left the main sequence by $z=0.4$, while 90 per cent of them are still part of the main sequence at $z=1$, and almost all (99 per cent) are active before $z=1.5$. The most massive passive galaxies with a $z=0$ stellar mass of $11.5\le\log (m_*/\Msun)<12.0$ have left the main sequence a long time ago. Around half of their progenitors have become passive by $z=1.3$, while 90 per cent are still active before $z=2$. Before $z=2.5$ all progenitors of the most massive galaxies are still on the main sequence. This shows that all passive galaxies have main progenitors on the main sequence at some redshift.

Having analysed the progenitors of passive galaxies at $z=0$ and determined when those galaxies become quenched, we now invert the question. Given the populations of galaxies that lie on the main sequence at a given redshift, we investigate how many of those active galaxies become quenched by $z=0$. For this we first select active galaxies that are on the main branch at a specified redshift and have a specified stellar mass. Instead of using stellar mass bins, we here employ stellar mass thresholds, to distinguish this analysis clearly from the previous one, as now the stellar mass is selected at a given (progenitor) redshift and not at the (descendant) redshift of zero. For each of the galaxies in the samples we then compute the fraction that has left the main branch and became passive by $z=0$. We show the resulting quenched fractions in Figure \ref{fig:descendantFQ} for four stellar mass thresholds. For the redshift indicated in the abscissa, the ordinate gives the fraction of all galaxies that are more massive than the threshold mass at this redshift, which will leave the main sequence by $z=0$.

More than 90 per cent of all galaxies on the main sequence at $z=2$ that are more massive than $\log (m_*/\Msun)=9.5$ will leave the main sequence by the present day. Even at a lower redshift of $z=1$, more than 85 per cent of all galaxies on the main sequence with $\log (m_*/\Msun)=9.5$ will be passive at $z=0$. More massive galaxies tend to have an even higher probability of being quenched and leaving the main sequence. Virtually all massive galaxies with $\log (m_*/\Msun)\ge11$ on the main sequence at $z=1$ will are passive today. Even at $z=0.5$, 95 per cent of all active massive galaxies will be quenched by $z=0$. At low redshift the values for the quenched descendant fractions drop steeply. Still, almost independent of stellar mass, about half of all galaxies that are on the main sequence at $z=0.1$ will leave the main sequence by $z=0$. This  shows that most of the galaxies that are observed on the main sequence at intermediate to high redshift ($z\ge1$) do not stay on the main sequence throughout their lives, but instead will eventually become passive and settle in the red cloud by $z=0$. This again confirms that the massive galaxies that are observed to have very high SFRs at $z=1-2$ are the progenitors of present-day passive ETGs.


\subsection{The progenitor populations of ETGs} \label{sec:progenitors}

\begin{figure*}
    \includegraphics[width=0.85\fullwidth]{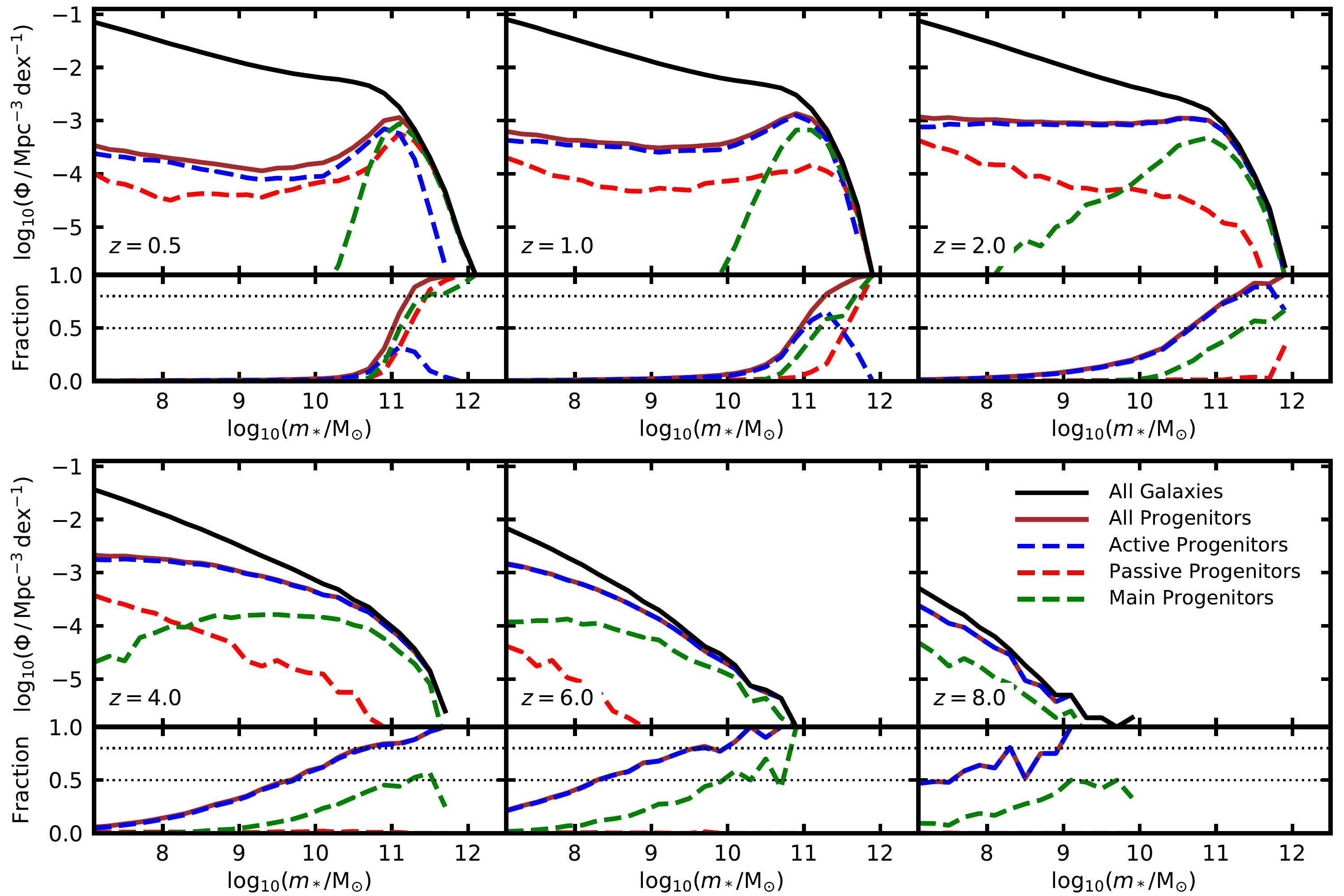}
    \caption{
    The SMFs of all progenitors of present quenched galaxies with $\log(m_*/\Msun) > 11$ at different redshifts (upper panels). The blue and red lines show the SMFs of all active and passive progenitors, respectively, the brown lines show the total SMFs of all progenitors and the green lines show the SMFs of the main progenitors. In comparison, the black lines show the total SMFs of all galaxies. The bottom panels indicate the fraction that the progenitors of passive galaxies contribute to the total
    SMF at each redshift. The dotted horizontal lines give fractions of 80 and 50 per cent.
    }
    \label{fig:progenitorSMF11}
\end{figure*}

\begin{figure*}
    \includegraphics[width=0.85\fullwidth]{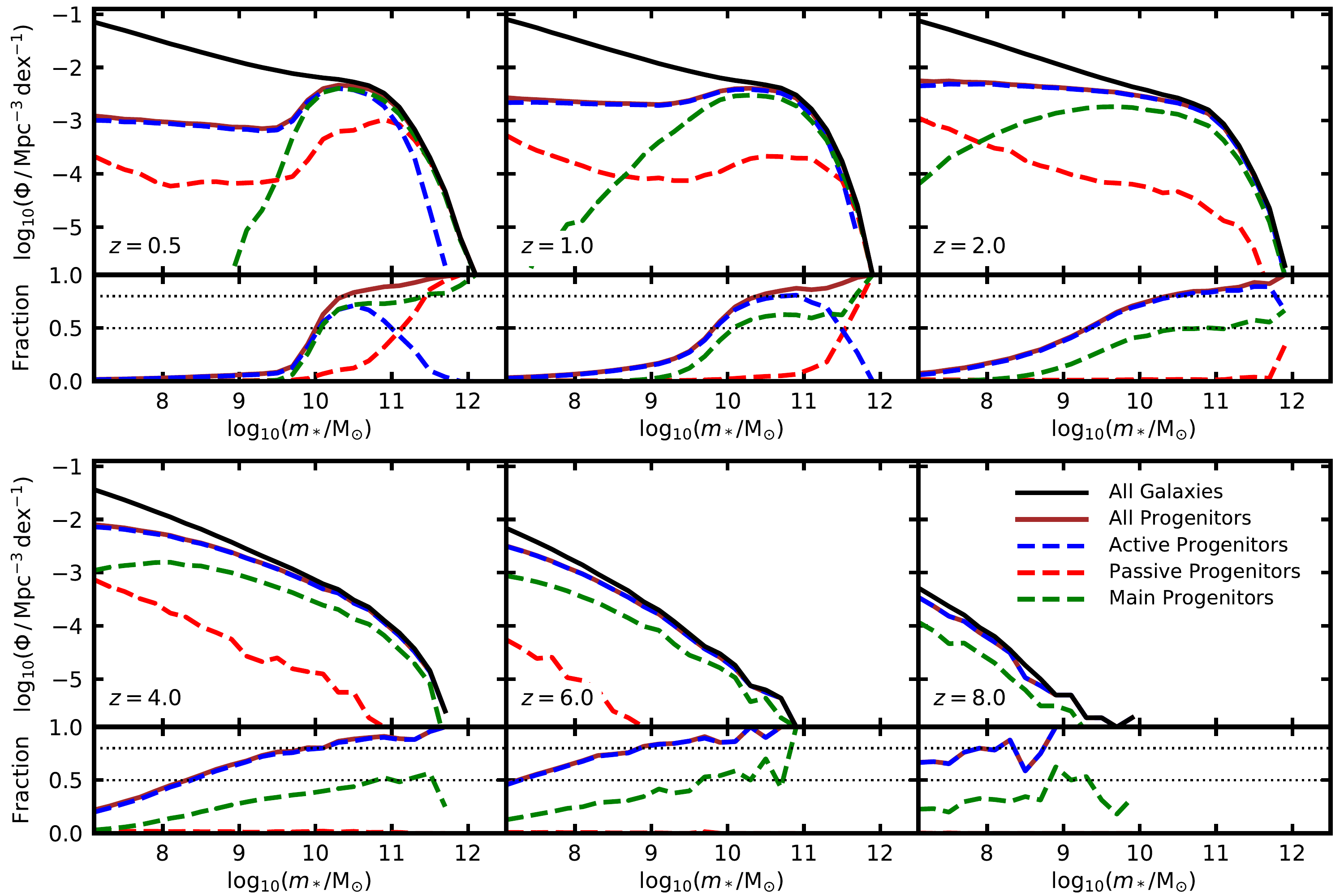}
    \caption{
    Same as Figure \ref{fig:progenitorSMF11}, but for progenitors of present quenched galaxies with $\log(m_*/\Msun) > 10$.
    }
    \label{fig:progenitorSMF10}
\end{figure*}

In the previous section we focused on the properties of the main progenitors of present ETGs, i.e. we traced the most massive progenitors in the merger trees through cosmic time. In this section we investigate the full distribution of the progenitors including the smaller galaxies that merged onto the main branch of the merger tree to study which galaxies at high redshift have accumulated to form the present population of passive galaxies. We further analyse how frequent the ETG progenitors are at high redshift, and how much of the total galaxy population they account for. First we select all passive galaxies at $z=0$ and create two samples: one massive sample in which all galaxies are more massive than $\log(m_*/\Msun) = 11$, and a sample that includes all galaxies that are more massive than $\log(m_*/\Msun) = 10$. We then trace the merger tress of all galaxies in these samples to high redshift and record all progenitors including the main progenitors and the progenitors that have merged. Using the populations at six redshifts, we determine their SMF. We also divide these populations into active and passive progenitors (with our standard specific SFR threshold) and determine their respective SMFs. Finally, we only select the main progenitors (including active and passive) and determine their SMF.

The resulting SMFs are shown in Figure \ref{fig:progenitorSMF11} for progenitors of passive galaxies with a $z=0$ stellar mass of $\log(m_*/\Msun) > 11$, and in Figure \ref{fig:progenitorSMF10} for progenitors of passive galaxies with a $z=0$ stellar mass of $\log(m_*/\Msun) > 10$. In both Figures, the black solid lines show the total SMFs of all galaxies at each redshift. The brown solid lines show the abundance of all progenitors of the $z=0$ samples, while the dashed blue and red lines divide this into components for active and passive progenitors, respectively. The green lines indicate the results obtained when only considering the main (most massive) progenitors of each $z=0$ galaxy. The top panels show the SMFs at six redshifts, while the bottom panel shows the relative fraction of each population, i.e. their contribution to the total SMF at this redshift. The dotted horizontal lines in the lower panels indicate fractions of 80 and 50 per cent of the total abundance of galaxies.

For both samples, we find that the progenitors of presently passive galaxies account for the massive end of the SMF. At low redshift ($z=0.5$) they contribute 50 per cent to the total SMF above the selected threshold mass, and over 80 per cent above slightly higher stellar masses of $\log(m_*/\Msun) = 10.1$ and $11.1$, respectively. At high redshift, the massive end consists almost exclusively of the progenitors of passive galaxies. At $z=4$ the progenitors of the massive sample account for over 80 per cent of the SMF above a mass of $\log(m_*/\Msun) = 10.5$, while at $z=8$ the ETG progenitors account for over 80 per cent of the SMF above $\log(m_*/\Msun) = 8$. For the sample with a lower $z=0$ threshold mass, we find that at $z=4$ more than 80 per cent of the total SMF above $\log(m_*/\Msun) = 9.5$ consists of the progenitors of this sample, and at $z=8$ more than 80 per cent of the SMF is made from the progenitors.This indicates that most of the galaxies that contribute to the SMFs found in present surveys can be accounted for by the progenitors of present-day ETGs with $\log(m_*/\Msun) > 10$.

The low-mass end of SMFs at low to intermediate redshift ($z\le4$) contains very few progenitors of passive galaxies. The progenitors of the massive sample contribute less than one per cent to the total SMF below $\log(m_*/\Msun) = 10$ at $z=0.5$, and less than 10 per cent below $\log(m_*/\Msun) = 10$ at $z=2$. The progenitors of the sample that include intermediate-mass galaxies, the contribution to the low-mass end of the SMFs is also negligible, although slightly higher than for the massive sample. At $z=0.5$ these progenitors contribute up to a few per cent below $\log(m_*/\Msun) = 10$, and at $z=2$ they contribute about 50 per cent at $\log(m_*/\Msun) = 9.5$ and 10 per cent at $\log(m_*/\Msun) = 8$. At high redshift the progenitors of presently passive galaxies significantly contribute to the total SMFs. The progenitors of the massive sample make up 50 per cent of the $z=4$ SMF at $\log(m_*/\Msun) = 9.5$, and 50 per cent of the $z=8$ SMF at $\log(m_*/\Msun) = 7$. Similarly, at $z=4$, 50 per cent of the SMF at $\log(m_*/\Msun) = 8$ consists of the progenitors of the sample with a lower threshold mass. This shows that at low redshift the progenitors of presently passive galaxies contribute little to the low-mass end of the observed SMFs, while a significant fraction of the observed low-mass galaxies at high redshift end up in present-day ETGs.

We further find that the progenitors of both samples at low redshift are mostly active at the low-mass end, and mostly passive at the massive end. At $z=0.5$ the majority of progenitors that are more massive than $\log(m_*/\Msun) = 11$ are passive, while below this mass most progenitors are actively forming stars. The contribution of passive galaxies diminishes at higher redshift and is insignificant above $z\gtrsim2$, i.e. almost all progenitors are forming stars at high redshift. For the contribution of the main progenitors to the total SMF, we find that at low redshift the main progenitors only account for the massive end of the SMF. The small contribution of low-mass progenitors to the SMF exclusively consists of galaxies that will merge onto a larger galaxy later. However, at high redshift we can identify a number of low-mass galaxies that are the main progenitors of presently passive galaxies. While there are more secondary progenitors than main progenitors at the low-mass end, roughly half of the progenitors at the massive end at high redshift are main progenitors. Moreover, they can contribute up to 50 per cent of the galaxies in the SMF at the massive end. Our analysis highlights that a large fraction of the observed galaxies at high redshift will eventually be accreted by a larger galaxy that is passive by $z=0$.


\section{Conclusions} \label{sec:conclusions}

We analysed the formation and evolution of ETGs with the empirical model {\sc emerge}, which follows the evolution of galaxies in individual dark matter haloes through cosmic time. For each halo in a cosmological $N$-body simulation, we computed the growth rate and calculate the SFR of the galaxy within the halo as the product of the halo growth rate, which specifies how much material becomes available, and the instantaneous baryon conversion efficiency, which specifies how efficiently this material can be converted into stars. Within \textsc{emerge}, the latter summarises all of the baryonic physics of galaxy formation and depends on halo mass and redshift. When a halo stops growing, the SFR is left constant for a specified amount of time and is then rapidly quenched. Galaxies in haloes that have lost a significant fraction of their mass are stripped. Once satellites lose their kinetic energy due to dynamic friction, they merge with the central galaxy and eject a fraction of their stars to the halo. We integrated the star formation history of each galaxy taking into account mass loss from dying stars to derive its stellar mass.

The model is constrained by a variety of observational data, such as SMFs, specific SFRs, the cosmic SFR density, fractions of quenched galaxies and clustering as a function of stellar mass. All model parameters were determined using an MCMC ensemble sampler by requiring that all observed data be reproduced simultaneously. This allowed us to connect the observations at different epochs and to follow the growth of stellar mass in galaxies as constrained by the observations. For each individual galaxy we determined if it was active or passive using a specific SFR threshold that corresponds to 30 per cent of the Hubble parameter at the observed redshift. The conclusions we can draw from \textsc{emerge} are therefore a direct consequence of the evolution of the observational data used to constrain the model over cosmic time in a $\Lambda$CDM Universe. As a result, we expect similar conclusions for any $\Lambda$CDM-based models, which fit the same observational data, e.g. empirical models such as the UniverseMachine \citep{Behroozi:2019aa}, or state-of-the-art semi-analytical models such as the model by \citet{Henriques:2019aa}. We expect somewhat more discrepant results from current hydrodynamic simulations since these typically fit the data somewhat less well, although future simulations that are in better agreement with observations are likely to derive similar results. In our analysis, we focused on ETG properties and derived the following conclusions:

(i) \textit{The stellar-to-halo mass relation in ETGs}: We found that at fixed halo mass, passive galaxies have a higher stellar mass, as they preferentially formed at higher redshift when the instantaneous conversion efficiency was higher. However, we also found that this relation cannot be simply inverted, and that at fixed stellar mass, ETGs live in more massive haloes than their LTG counterparts. The reason for this effect is the scatter in the SHM relation, which causes many active galaxies in haloes of lower mass to have a high stellar mass. In massive haloes there are fewer active galaxies with higher stellar mass, so that the average halo mass at fixed stellar mass is biased towards lower masses for active galaxies. We compared our results to the observational weak lensing constraints by \citet{Mandelbaum:2006aa, Mandelbaum:2016aa} and found excellent agreement.

(ii) \textit{The growth of ETGs through accreted satellites}: We calculated the fraction of the stellar mass that has formed ex-situ for each galaxy and determined the average ex-situ fraction as a function of halo and stellar mass. Because of the shape of the SHM ratio low-mass haloes typically accrete satellites that are much less massive then the central galaxy, such that the ex-situ fraction is low. However, in massive haloes the conversion efficiency is suppressed, such that the stellar mass of accreted satellites is of the order of the central galaxy's mass, resulting in a high ex-situ fraction. When taking into account the mass of the central galaxy and the mass in the ICM, we found that ex-situ star formation starts to become relevant for haloes with $\log M_\mathrm{h}/\Msun\gtrsim12$ and reaches 50 per cent around $\log M_\mathrm{h}/\Msun\sim13$. For the most massive haloes, the ex-situ fraction approaches unity. Taking only into account the stellar mass in the central galaxy (i.e. excluding the ICM), we found lower ex-situ fractions that reach 50 per cent around $\log M_\mathrm{h}/\Msun\sim13.5$ and 80 per cent for the most massive haloes. Comparing our results to other models, we found that \textsc{emerge} predicts a lower ex-situ fraction for intermediate-mass haloes with $\log M_\mathrm{h}/\Msun\sim13-14$. Finally, we found that the ex-situ fraction is a strong function of stellar mass. Below $\log(m_*/\Msun) = 11$ the ex-situ fraction is negligible, but then rises quickly to 60 per cent at $\log(m_*/\Msun) = 11.5$, in agreement with \citet{Bernardi:2019aa}. The transition mass for which accreted stellar mass becomes relevant is only a weak function of redshift.

(iii) \textit{Star formation and mass assembly in ETGs}: Tracing the stellar mass and SFRs of the main progenitors of present-day galaxies through cosmic time, we found that low-mass galaxies tend to assemble late although their dark matter haloes typically assemble at high redshift. In contrast, massive galaxies tend to assemble early, while their dark matter haloes assemble late. This effect that has been called `anti-hierarchical' galaxy formation or `downsizing' can be easily explained invoking the conversion efficiency. While low-mass haloes are already in place at high redshift, their galaxies can convert gas into stars only very inefficiently until the halo becomes massive enough, delaying star formation in low-mass galaxies. In massive haloes, the galaxies can only grow efficiently at high redshift, before the halo mass becomes too high and only allows for low SFRs at late times. Only for the most massive galaxies we found relatively late assembly times. While the stars in those haloes formed at high redshift a large fraction gets accreted by the massive galaxy only at late times. Further, we found star formation histories that peak at higher redshift for more massive systems. For the most massive systems $\log(m_*/\Msun) > 11.5$, we found that most stars have formed before $z=3$, and over 90 per cent of the present stellar mass had formed by $z=1$. Our results are in agreement with estimates from IFU surveys derived through galactic archeology.

(iv) \textit{The build-up of the star formation bimodality}: We derived the relation between stellar mass and SFR up to high redshift and found that the `main sequence of star formation' starts to form before $z=8$ and is in place by $z=4$ comprising the vast majority of galaxies. Around $z=2$, we found that the main sequence develops a bending at the massive end, as star formation in the most massive galaxies is becoming inefficient. We also found that this downturn can only be explained by a gradual decline of the conversion efficiency, i.e. feedback has to become stronger gradually, while a model where feedback quenches massive galaxies instantaneously does not result in a bending. The `red sequence' is beginning to emerge at $z\sim2$ and grows until $z=0$, when star formation in massive galaxies has become very inefficient. Our results are in good agreement with major surveys. We further investigated when present-day ETGs are quenched, and found that low-mass galaxies become passive very late, while massive galaxies finally stop forming stars by $z=1$. Above this redshift, most ETG main progenitors are located on the main sequence. Conversely, more than 90 per cent of the galaxies on the main sequence at $z=2$ with $\log(m_*/\Msun) > 10$ evolve into present-day ETGs.

(v) \textit{The progenitor populations of ETGs}: Following the full distribution of ETG progenitors including galaxies that merge onto them, we derived the SMF of all ETG progenitors, as well as active and passive ETG progenitors. We found that the total ETG progenitor population accounts for the massive end of the SMF up to high redshift. At low redshift, the ETG progenitors contribute very little to the low-mass end of the SMF. However at high redshift, most of the galaxies in the SMF, even at low stellar mass, are the progenitors of ETGs. Above redshift 6, more than 80 per cent of the SMF above $\log(m_*/\Msun) = 9$ can be accounted for by progenitors of present-day ETGs with $\log(m_*/\Msun) > 10$. In contrast, we found that the main progenitors of those ETGs only account for less than 50 per cent of the $z=6$ SMF above $\log(m_*/\Msun) = 9$.

Our final conclusion is that current and future high redshift observations mainly probe the formation of present-day ETGs.


\section*{Acknowledgements}

We thank
Yohan Dubois,
Bruno Henriques,
Richard McDermid,
Rhea-Silvia Remus,
Annalisa Pillepich,
and Daniel Thomas
for sharing their data.
We are also grateful to 
Peter Behroozi,
Andreas Burkert,
Michael Fall,
Andrew Hearin,
Andrey Kravtsov,
Aura Obreja,
Joseph O'Leary,
Jerry Ostriker,
Ulrich Steinwandel,
and
Frank van den Bosch
for enlightening discussions.
The cosmological simulations used in this work were carried out at the Freya Cluster at the Max Planck Computing and Data Facility in Garching, and the Leibniz Supercomputing Centre.
BPM acknowledges an Emmy Noether grant funded by the Deutsche Forschungsgemeinschaft (DFG, German Research Foundation) -- MO 2979/1-1.




\bibliographystyle{mnras} \bibliography{astro}

\bsp	
\label{lastpage}
\end{document}